\RequirePackage{fix-cm}
\documentclass[twocolumn,epjc3]{svjour3}  

\smartqed  
\RequirePackage{graphicx}
\journalname{Eur. Phys. J. A}
\usepackage{xspace}
\usepackage{cite}
\usepackage{stfloats}
\usepackage{color}

\newcommand{\pp}    {pp\xspace}

\newcommand{\PbPb}  {\mbox{Pb--Pb}\xspace}

\newcommand{\s}            {\ensuremath{\sqrt{s}}\xspace}

\newcommand{\Pt}           {\ensuremath{p_\mathrm{T}}\xspace}

\newcommand{\dndeta}       {\ensuremath{\mathrm{d}N_\mathrm{ch}/\mathrm{d}\eta}\xspace}
\newcommand{\avdndeta}     {\ensuremath{\langle\dndeta\rangle}\xspace}

\newcommand{\nineH}         {$\sqrt{s}~=~0.9$~Te\kern-.1emV\xspace}
\newcommand{\seven}         {$\sqrt{s}~=~7$~Te\kern-.1emV\xspace}
\newcommand{\twoH}          {$\sqrt{s}~=~0.2$~Te\kern-.1emV\xspace}
\newcommand{\twosevensix}   {$\sqrt{s}~=~2.76$~Te\kern-.1emV\xspace}
\newcommand{\five}          {$\sqrt{s}~=~5.02$~Te\kern-.1emV\xspace}
\newcommand{\twosevensixnn} {$\sqrt{s_{\mathrm{NN}}}~=~2.76$~Te\kern-.1emV\xspace}
\newcommand{\fivenn}        {$\sqrt{s_{\mathrm{NN}}}~=~5.02$~Te\kern-.1emV\xspace}

\newcommand{\GeVc}          {Ge\kern-.1emV/$c$\xspace}
\newcommand{\MeVc}          {Me\kern-.1emV/$c$\xspace}
\newcommand{\TeV}           {Te\kern-.1emV\xspace}
\newcommand{\GeV}           {Ge\kern-.1emV\xspace}
\newcommand{\MeV}           {Me\kern-.1emV\xspace}
\newcommand{\GeVmass}       {Ge\kern-.2emV/$c^2$\xspace}
\newcommand{\MeVmass}       {Me\kern-.2emV/$c^2$\xspace}

\newcommand{\kzero}  {\ensuremath{{\rm K}^{0}_{\rm{S}}}\xspace}
\newcommand{\kstar}  {\ensuremath{{\rm K}^{*0}}\xspace}
\newcommand{\lmb}    {\ensuremath{{\rm\Lambda}}\xspace}
\newcommand{\almb}   {\ensuremath{\overline{\lmb}}\xspace}
\newcommand{\Om}     {\ensuremath{{\rm\Omega^-}}\xspace}
\newcommand{\Mo}     {\ensuremath{\overline{{\rm\Omega}}^+}\xspace}
\newcommand{\X}      {\ensuremath{{\rm\Xi^-}}\xspace}
\newcommand{\Ix}     {\ensuremath{{\rm\overline{\Xi}}^+}\xspace}
\newcommand{\Xis}    {\ensuremath{{\rm\Xi}^{\pm}}\xspace}
\newcommand{\XI}{\ensuremath{{\rm\Xi}}\xspace}
\newcommand{\OMS}{\ensuremath{{\rm\Omega}}\xspace}
\newcommand{\Oms}    {\ensuremath{{\rm\Omega}^{\pm}}\xspace}
\newcommand{\PHI}    {\ensuremath{{\rm\phi}}\xspace}

\newcommand{\pT}{\ensuremath{\Pt}\xspace}
\newcommand{\pTj}{\ensuremath{p_{\rm T,~jet}}}
\newcommand{\pTjch}{\ensuremath{\pTj^{\rm ch}}}

\newcommand{\kT}{\ensuremath{k_{\rm T}}}
\newcommand{\akT}{anti-\kT\xspace}
\newcommand{\thirteen} {$\sqrt{s}~=~13$~Te\kern-.1emV\xspace}

\newcommand{\toprule}{\noalign{\smallskip}\hline}
\newcommand{\midrule}{\noalign{\smallskip}\hline\noalign{\smallskip}}
\newcommand{\bottomrule}{\noalign{\smallskip}\hline}
\begin{document}
\begin{sloppypar}
\title{Strange particle production in jets and underlying events in \pp collisions at \seven with PYTHIA8 generator
}


\author{Pengyao Cui\thanksref{e1, addr1}
        \and
        Zhongbao Yin\thanksref{e2, addr1}
        \and
        Liang Zheng\thanksref{e3, addr2, addr1}
}

\thankstext{e1}{e-mail:cuipengyao@mails.ccnu.edu.cn}
\thankstext{e2}{e-mail:zbyin@mail.ccnu.edu.cn}
\thankstext{e3}{e-mail:zhengliang@cug.edu.cn (corresponding author)}


\institute{
Key Laboratory of Quark and Lepton Physics (MOE) and Institute of Particle Physics,
Central China Normal University,
Wuhan 430079, China \label{addr1}
   \and
School of Mathematics and Physics, China University of Geosciences (Wuhan), Wuhan 430074, China \label{addr2}
}

\date{Received: date / Accepted: date}

\maketitle

\begin{abstract}
Strange hadron production in pp collisions at \seven is studied in jets and underlying events using the PYTHIA8 event generator. Matching strange hadrons to the jet area and the underlying event area is expected to help us disentangle the strange particles produced in hard and soft processes. The yield and the relative production of strange hadrons dependent on the event multiplicity are investigated with the color reconnection and color rope mechanisms implemented in the PYTHIA8 framework. It is found that the inclusive strange hadron productions can be reasonably described by the color reconnection and color rope combined effects. A significant multiplicity dependent enhancement of the strange baryon production in the jet area is observed induced by the modified string fragmentation mechanisms, indicating the strange baryon enhancement persists in both the hard and the soft process.
Multi-strange baryons are found to be more collimated with the jet axis than other strange hadrons in the string fragmentation picture with the jet shape analysis technique. Future experimental examination of these jet related strange hadron productions will provide more insight to the origin of strangeness enhancement in small systems.

\end{abstract}

%


%
\section{Introduction}
\label{sec:intro}

In heavy-ion collisions at ultra-relativistic energies, it is well established that a new state of nuclear matter called Quark-Gluon-Plasma~(QGP) can be created~\cite{Adams:2005dq, Adcox:2004mh, Arsene:2004fa, Back:2004je, Schukraft:2011na, Satz:2000bn, Shuryak:1983ni, Jacak:2012dx, Cleymans:1985wb, Bass:1998vz, BraunMunzinger:2007zz}. Proton-proton (\pp), proton-nucleus (p--A) collisions have been suggested to serve as the references to study heavy-ion collisions and quantify the properties of the QGP. However, recent measurements have revealed strong flow-like effects in high multiplicity \pp and p--A collisions at LHC energies which are qualitatively similar to that observed in heavy-ion collisions~\cite{Abelev:2012sk, Chatrchyan:2013eya, Khachatryan:2010gv, CMS:2012qk, Abelev:2012ola, Aad:2012gla, Aad:2013fja, Chatrchyan:2013nka, Adare:2013esx, Adams:2006nd, Aad:2015gqa, ABELEV:2013wsa, Khachatryan:2015waa, Acharya:2019vdf, Abelev:2014uua, Adam:2015vsf}.
These include the long-range two particle angular correlations~\cite{Aad:2015gqa, Abelev:2012ola, ABELEV:2013wsa}, non-vanishing 2$^\mathrm{nd}$ order Fourier coefficients ($v_{2}$) in multi-particle cumulant studies~\cite{Acharya:2019vdf, Khachatryan:2015waa}, the baryon-to-meson enhancement at intermediate transverse momentum ($\pT$) range~\cite{ALICE:2018pal} and the strangeness enhancement effects~\cite{ALICE:2016fzo}. Therefore, it is of great interest to investigate the origin of the collectivity-driven features in small systems both experimentally and theoretically.

In particular, the strangeness enhancement, referring to the enhanced production of multi-strange baryons, has been historically regarded as an important signature of the QGP formation. For hadronic collisions in small systems like \pp and p--A collisions, the production of particles containing strange quarks are supposed to be suppressed due to the large strange quark mass. The chemically equilibrated QGP medium, on the other hand, may create a large number of strange quarks via the thermal gluon fusion process and enhance the formation of multi-strange hadrons. It is therefore a striking observation that a universal multi-strange hadron enhancement feature is reported by the ALICE collaboration~\cite{ALICE:2016fzo,ALICE:2018pal} across collision systems from \pp, p--A to AA collisions. The measured multiplicity dependent strange (\kzero, \lmb and \almb) and multi-strange (\X, \Ix, \Om and \Mo) particle production in \pp collisions at \seven is qualitatively similar to that observed in \PbPb collisions and cannot be easily reproduced by the conventional string fragmentation model. It is found that the improvements considering the inter-string interactions like color reconnection or color rope mechanisms are needed to describe these novel flavor dependent observables~\cite{Nayak:2018xip, Bierlich:2016vgw, OrtizVelasquez:2013ofg}. 

In this paper, we will study the production of strange hadrons, \kzero,  \lmb, \Xis, \Oms and the resonance states \kstar and \PHI within the jet and the underlying event region which separates the contribution associated with hard and soft processes. The rest of the paper is organized as follows: in Sec.~\ref{sec:model}, we give a short introduction about the modified string fragmentation models implemented in the PYTHIA8 framework. The analysis method involving jet observables is described in Sec.~\ref{sec:ana}. We compare the model calculations with the experimental data and make predictions for the in-jet and underlying event strange hadron productions in Sec.~\ref{sec:result}. In the end, we come to our summary in Sec.~\ref{sec:sum}. 

\section{Color reconnection and color rope models}
\label{sec:model}
In this work, we perform the study using the PYTHIA8 (version 8.2.4) event generator together with the hadronization models considering color reconnection and color rope mechanisms. Within the multiple parton interaction (MPI) framework~\cite{AxialFieldSpectrometer:1986dfj,UA2:1991apc,CDF:1997yfa} implemented in PYTHIA8, a large number of final state string objects can be generated over limited transverse space in high energy collisions. The traditional string fragmentation process is expected to be modified in the high parton density environment due to the inter-string interactions. Based on the Monash 2013 tune, color reconnection~\cite{Christiansen:2015yqa,Sjostrand:2014zea} and color rope~\cite{Biro:1984cf,Bialas:1984ye} effects will be included to investigate strange hadron production from both hard processes related to the jet production and soft processes in underlying events. 

\paragraph{Color reconnection model}
Color reconnection (CR) deals with the formation of strings by connecting different final state partons via color lines in a way that the total string length is as short as possible. Within the MPI framework, partons from independent hard scatterings at mid-rapidity can reconnect and make a significant transverse boost to the created string piece due to the involvement of more mid-rapidity partons~\cite{OrtizVelasquez:2013ofg}. Different CR models have been developed based on the calculation of the probability to connect partons by incorporating various color flow structures. In this work, we employ the beyond leading color (BLC) CR model, in which strings are allowed to form between both leading and non-leading connected partons~\cite{Christiansen:2015yqa}. As the CR model connects both low and high $\pT$ partons, both soft and hard processes will be affected by this mechanism. With the possibility to form junction in beyond leading color CR as additional source for baryon production, a multiplicity dependent baryon enhancement is observed in this model~\cite{Bierlich:2015rha}. The parameters of BLC mode used in this paper are listed in Table~\ref{tab:CRparameter} with respect to the default settings of the Monash Tune. 
\begin{table}[ht]
    \caption{Parameters of Beyond Leading Color reconnection model with respect to the default settings of the Monash Tune}
    \resizebox{0.48\textwidth}{!}{
      \begin{tabular}{lll}
       \toprule
	    Parameters & Monash & BLC  \\
	    \midrule 
	    MultiPartonInteractions:pT0Ref &  2.28 & 2.15\\ 
	    BeamRemnants:remnantMode & 0 & 1 \\
	    BeamRemnants:saturation & -& 5 \\
	    ColourReconnection:reconnect &on& on \\
	    ColourReconnection:mode & 0 & 1 \\
	    ColourReconnection:allowDoubleJunRem & on&off \\
	    ColourReconnection:m0 & - & 0.3  \\
	    ColourReconnection:allowJunctions & - & on \\
	    ColourReconnection:junctionCorrection & - & 1.2 \\
	    ColorReconnection:timeDilationMode & - & 2 \\
	    ColourReconnection:timeDilationPar & - & 0.18\\ 
	    \bottomrule   
      \end{tabular}
    }
    \label{tab:CRparameter}
\end{table}

\paragraph{Color rope model} The color rope hadronization model is constructed based on the idea that strings overlapped with each other in geometric space may act coherently to form a color rope~\cite{Flensburg:2011kk}. Each string piece is treated as a flux tube with a model dependent transverse size, which allows to calculate the amount of overlap between strings with a given impact parameter of the event. 
The interfered strings ended up forming ropes are hadronized with higher effective string tensions, reflecting the fact that more energy is available for the hadronization process. This effect can be implemented by connecting the relevant Lund string fragmentation parameters in PYTHIA to the effective string tension values following the Schwinger tunneling mechanism~\cite{Bierlich:2014xba, Zheng:2018yxq}. The breakup of strings with higher string tension in the dense environment is expected to produce more di-quarks and strange quarks, resulting in enhanced production of baryons and strange hadrons. 
The parameters of color rope model used in this paper are listed in the Table~\ref{tab:Ropeparameter}. 

\begin{table}[ht]
	\begin{center}
	  \caption{Color rope hadronization model parameters used in this study.}
		\begin{tabular}{lc}
			\toprule
			Parameters & Values \\
			\midrule
			Ropewalk:RopeHadronization & on \\
			Ropewalk:doShoving & on  \\
			Ropewalk:tInit & 1.5 \\
			Ropewalk:deltat & 0.05 \\
			Ropewalk:tShove & 0.1 \\
			Ropewalk:gAmplitude & 0. \\
			Ropewalk:doFlavour & on \\
			Ropewalk:r0 & 0.5 \\
			Ropewalk:m0 & 0.2 \\
			Ropewalk:beta & 0.1 \\
			\bottomrule
		\end{tabular} 
	\end{center}
	\label{tab:Ropeparameter}
\end{table}

\section{Method}\label{sec:ana}
The analysis is done by generating around 2.5 billion non-diffractive events for \pp collisions at \s = 7 TeV with CR, Rope hadronization and combined CR and Rope hadronization (labeled as ``CR + Rope") processes, respectively. The Monash tune is used as a reference to quantify the effects of each model. The events are categorized into ten different event classes based on the final state charge multiplicities within the pseudorapidity range of 2 $<$ $|\eta|$ $<$ 5. The mean pseudorapidity density of charged particles, \avdndeta is estimated for $|\eta|$ $<$ 0.5. We present the \avdndeta in each event class from our simulation compared to that measured by ALICE~\cite{ALICE:2016fzo} in Table~\ref{tab:eventclass}.
\begin{table*}[ht]
	\begin{center}
		\caption{Definition of the event multiplicity classes as fractions of the analyzed event sample and their
			corresponding $\avdndeta_{|\eta|<0.5}$ within $|\eta_\mathrm{lab}| < 0.5$.}
		\label{tab:eventclass}
		\resizebox{\textwidth}{!}{
		\begin{tabular}{lcccccccccc}
			\toprule
			Event class & I & II & III & IV & V & VI & VII & VIII & IX & X \\
			\midrule
			$\sigma/\sigma_\mathrm{INEL > 0}$ & 0-0.95\% & 0.95-4.7\% & 4.7-9.5\% & 9.5-14\% & 14-19\% & 19-28\% & 28-38\% & 38-48\% & 48-68\% & 68-100\%\\
			Exp data & 21.3 $\pm$ 0.6 & 16.5 $\pm$ 0.5 & 13.5 $\pm$ 0.4 & 11.5 $\pm$ 0.3 & 10.1 $\pm$ 0.3  & 8.45 $\pm$ 0.25 & 6.72 $\pm$ 0.21 & 5.40 $\pm$ 0.17 & 3.90 $\pm$ 0.14 & 2.26 $\pm$ 0.12  \\
			Monash & 19.2& 15.6 & 13.1 & 11.4 & 10.0 & 8.4 & 6.7 & 5.1 & 3.5 & 2.2\\
			CR & 18.8 & 15.6 & 13.3 & 11.5 & 10.3  & 8.6 & 6.7 & 5.4 & 3.8 & 2.2 \\
			Rope & 20.3 & 16.6 & 13.9 & 12.2 & 10.8 & 9.2 & 7.5 & 5.9 & 4.3 & 2.6 \\
			CR + Rope & 18.3 & 15.2 & 12.9 & 11.4 & 10.0 & 8.5 & 6.8 & 5.5 & 4.1 & 2.5 \\
			\bottomrule
		\end{tabular}
		}
	\end{center}
\end{table*}

We study the strange hadron productions from soft and hard processes following the same analysis procedure done in the ALICE experiment~\cite{ALICE:2021cvd} with the help of jet observables.
The strange hadrons considered in this work are usually reconstructed via their charged decay channels in the experiment. In our simulation, we turn off their decays in PYTHIA8 and calculate the yield of each strange hadron after considering the realistic experimental acceptance.
We use the charged-particle jet reconstructed using the \akT algorithm~\cite{Cacciari:2008gp} from the FastJet package~\cite{Cacciari:2011ma, Cacciari:2005hq}. The \akT algorithm is the most widely used jet finding algorithm in heavy ion physics, providing a much faster replacement compared to the older cone algorithms and removing the irregular jet boundaries existed in the \kT algorithm~\cite{Cacciari:2008gp}. We use the jet resolution parameter $R = 0.4$ to match the ALICE experimental condition in a wide range of jet related studies~\cite{ALICE:2021cvd,ALICE:2018ype,ALICE:2017svf}. A transverse momentum cut of $\pTj > 10$~\GeVc has been applied to the reconstructed charged-particle jets to select the jet events~\cite{ALICE:2021cvd}. The strange hadron production associated with hard processes is explored by investigating the particles produced inside the jet cone. To ensure that the in-jet particle acceptance is fully overlapping with the acceptance of strange particles ($|\eta_\mathrm{par}| < 0.75$), the jet axis cut is set to $|\eta_\mathrm{jet}| < 0.35~(0.75 - 0.4)$. Particles with distances to the jet axis $R(\mathrm{p, ~jet})$ less than a cone size $R_{\rm cone} = 0.4$ are defined as the in-jet particles, in which $R(\mathrm{p, ~jet})$ is calculated in the pseudorapidity and azimuthal angle space~($\eta - \varphi$ plane) as follows 
$$
	R(\mathrm{p, ~jet}) = \sqrt{\left( \eta_\mathrm{jet} - \eta_\mathrm{par} \right)^{2} + \left( \varphi_\mathrm{jet} - \varphi_\mathrm{par} \right)^{2}}.
$$

\begin{figure}[ht]
	\begin{center}
		\includegraphics[width=.3\textwidth]{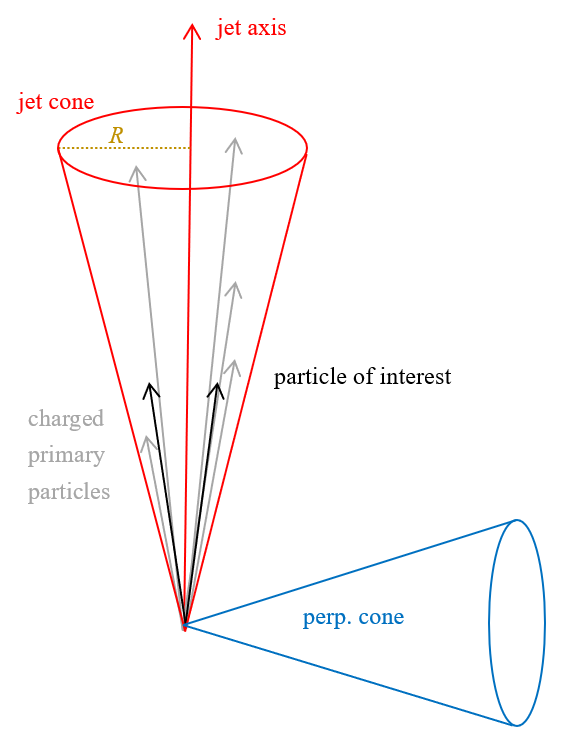}
	\end{center}
	\caption{Diagram illustrating the particle in jet cone region ($R$ (p, jet) $<$ 0.4) and perpendicular cone region (UE).}
	\label{fig:carton}
\end{figure}

The background contribution to in-jet particles from the underlying event (UE), which refers to particles not associated with hard scatterings in jet $R$ (p, jet) $<$ 0.4 region, is estimated in the perpendicular cone (PC) to the jet axis with a radius $R_{\rm cone} = 0.4$, as shown in Fig.~\ref{fig:carton}.

To subtract the UE component from the jet cone selection, a particle density per unit area $\rho$ is defined as
$$
   \rho = N_\mathrm{par}/(N_\mathrm{jet, ev}A_\mathrm{jet}),
$$
where $N_\mathrm{par}$ is integrated yields of strange hadrons located in the jet cone or PC, $N_\mathrm{jet, ev}$ is the number of generated events containing at least one selected charged-particle jets, $A_\mathrm{jet}$ is the area of all jet cones or PC occupied area in $\eta - \varphi$ plane in an event. Note that the inclusive strange hadron distributions are usually obtained with $|y| < 0.5$, slightly different from that of the in-jet particle acceptance $|\eta_\mathrm{par}| < 0.75$. 

\section{Results}\label{sec:result}
\begin{figure}[ht]
	\begin{center}
		\includegraphics[width=.46\textwidth]{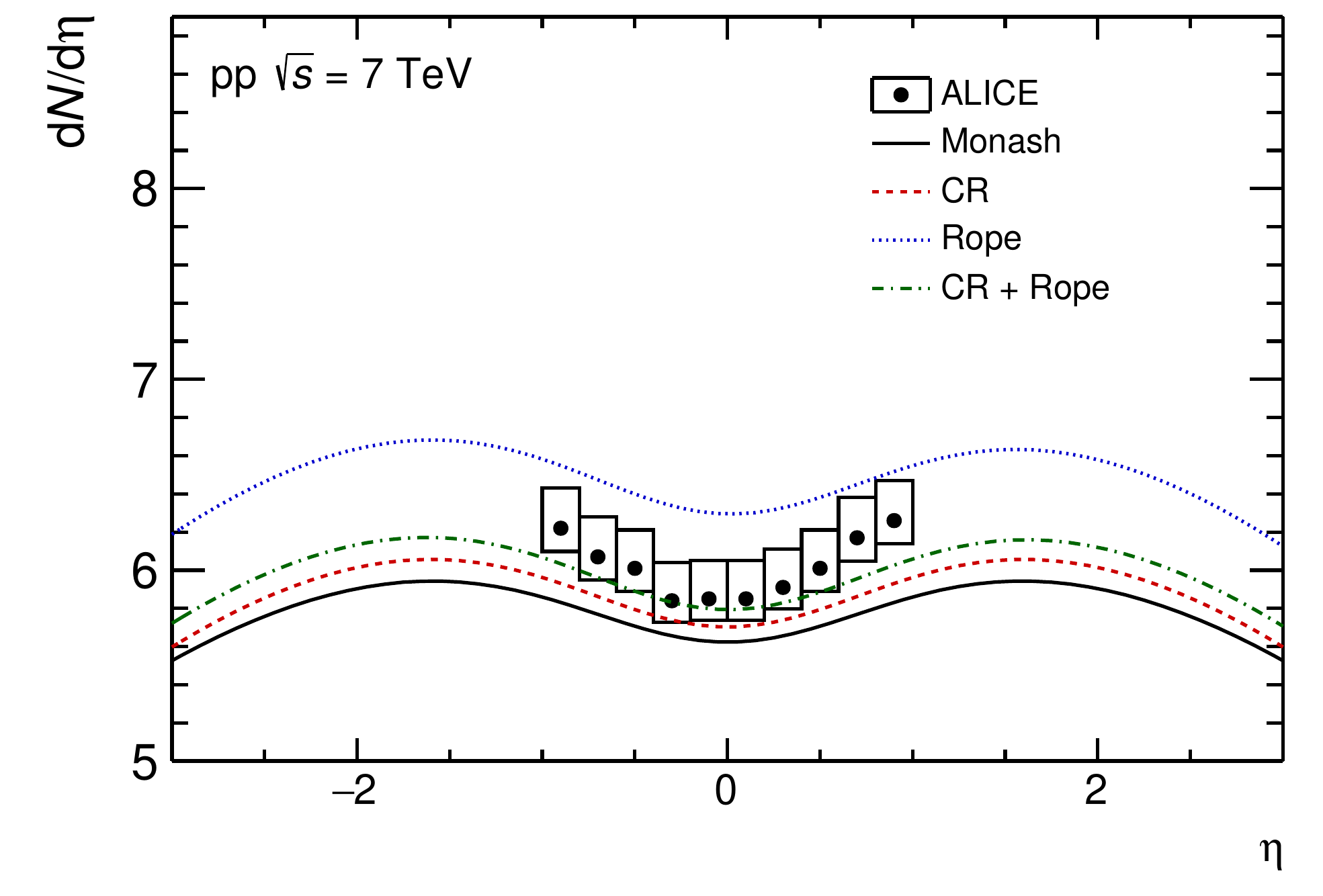}
		\includegraphics[width=.46\textwidth]{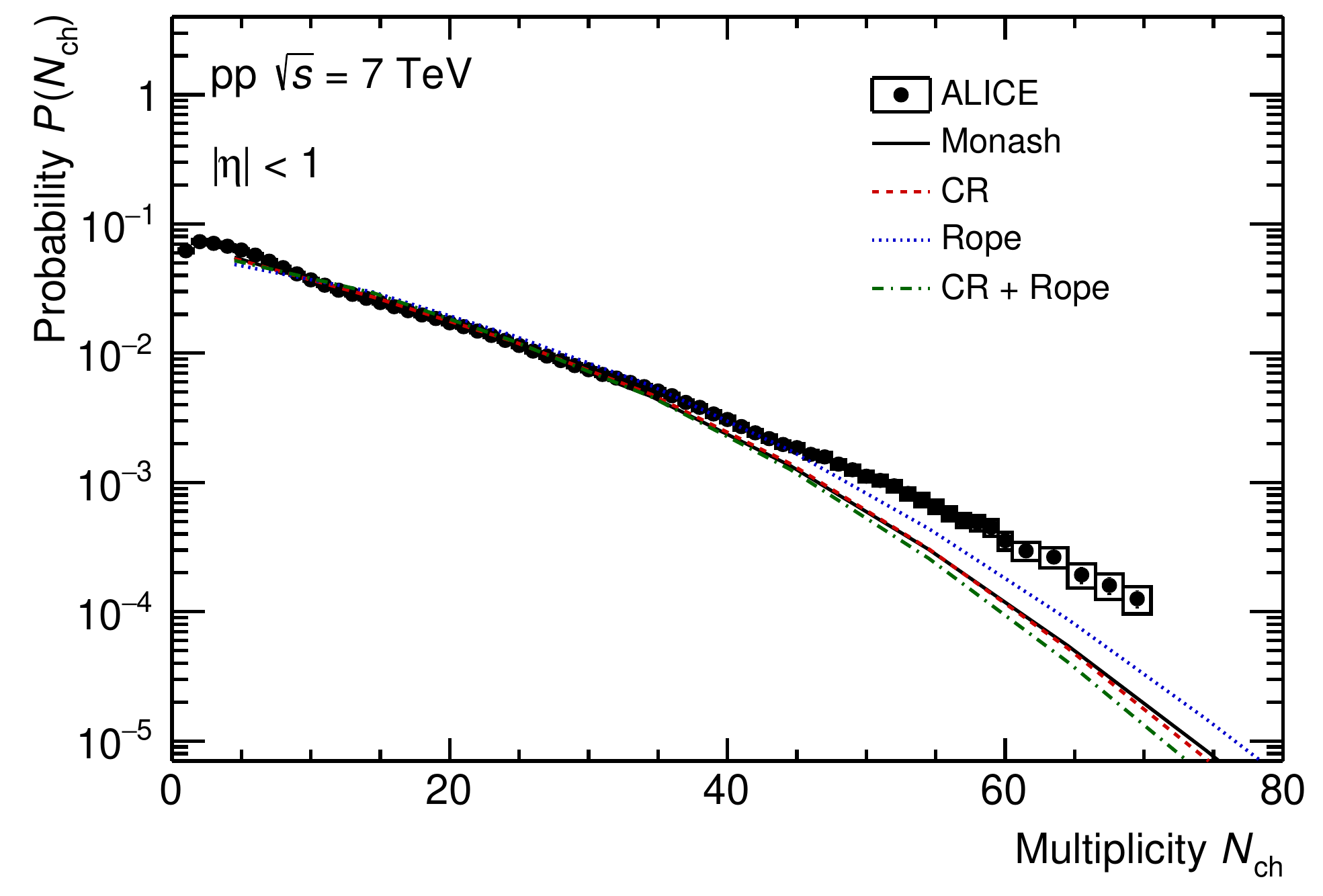}
	\end{center}
	\caption{Charged particle pseudo-rapidity distribution ($\dndeta$) (top) and multiplicity distribution (bottom) in \pp collisions at \seven. The experimental data are taken from~\cite{ALICE:2010mty}.}
	\label{fig:trkinfo}
\end{figure}
The results of \kzero, \kstar, \PHI, \lmb, \Xis and \Oms productions in \pp collisions at \seven from PYTHIA8 will be compared to the ALICE measurements in this section.
The inclusive charged particle pseudorapidity distribution ($\dndeta$) and the multiplicity distribution with color reconnection and color rope effects are compared with the corresponding experimental data in Fig.~\ref{fig:trkinfo}. The experimental data are shown in black point with the error box. The PYTHIA8 results with the Monash tune are shown in the black solid line. The CR and rope model effects are presented with the red dashed line and the blue dotted line, respectively. The green dash-dotted line includes both CR and rope effects. In the BLC CR model, reformulated strings may produce more heavy hadrons and hadrons with larger \pT, the multiplicity in this model is therefore supposed to be suppressed compared to the normal string fragmentation picture. Previous studies suggest that the MPI regulating parameter pT0Ref needs to be decreased to compensate this effect~\cite{Bierlich:2015rha}, as already seen in Tab.~\ref{tab:CRparameter}. We see in the comparison of Fig.~\ref{fig:trkinfo} (top) that the charged particle densities with the CR effect are consistent with the experimental data. It is also interesting to see that the charged particle yields are increased by about 10\% in the rope model with respect to the Monash tune. We also notice that the different fragmentation models can hardly reproduce the experimental data shown by Fig.~\ref{fig:trkinfo} (bottom) in the high multiplicity region. A more exclusive study to describe the charge particle distribution over a wide multiplicity range requires a systematic tuning to the relevant model parameters, which is beyond the scope of this work and should not change our conclusions on the strange hadron productions discussed in the rest of this paper. 

\subsection{Inclusive strange particle production}
\label{subsec:inclres}

\begin{figure*}[ht]
	\begin{center}
		\includegraphics[width=.32\textwidth]{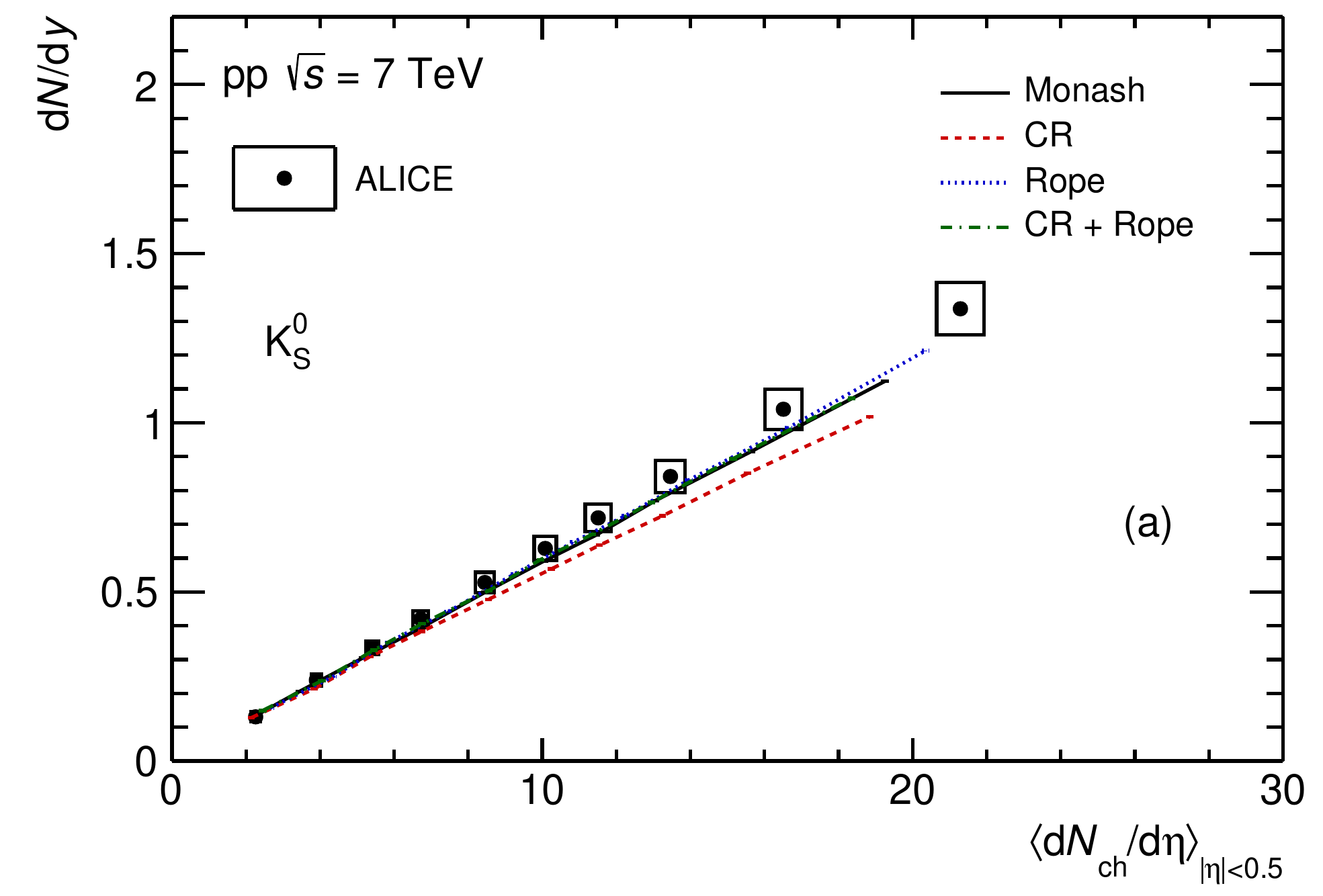}
		\includegraphics[width=.32\textwidth]{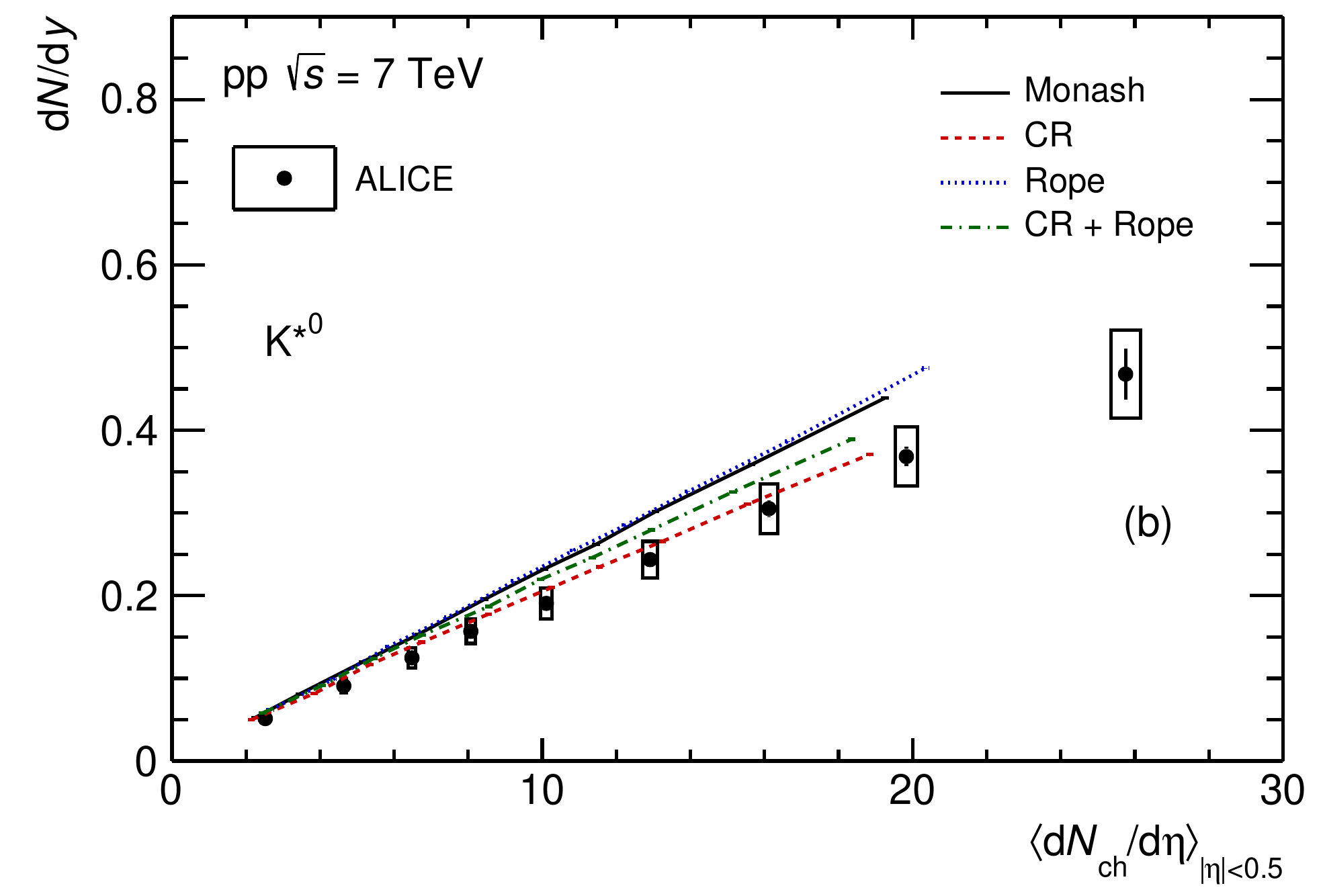}
		\includegraphics[width=.32\textwidth]{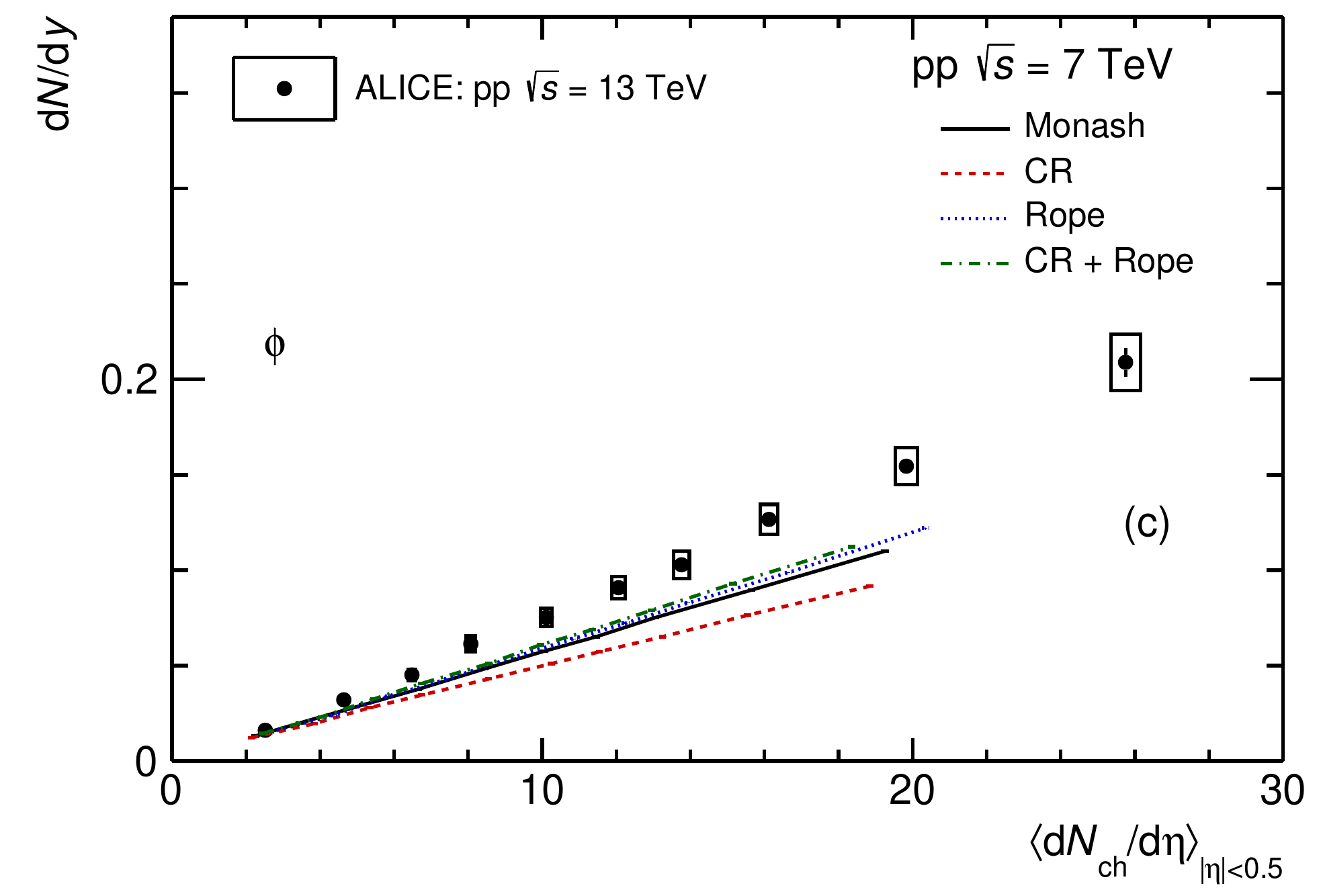}
		\includegraphics[width=.32\textwidth]{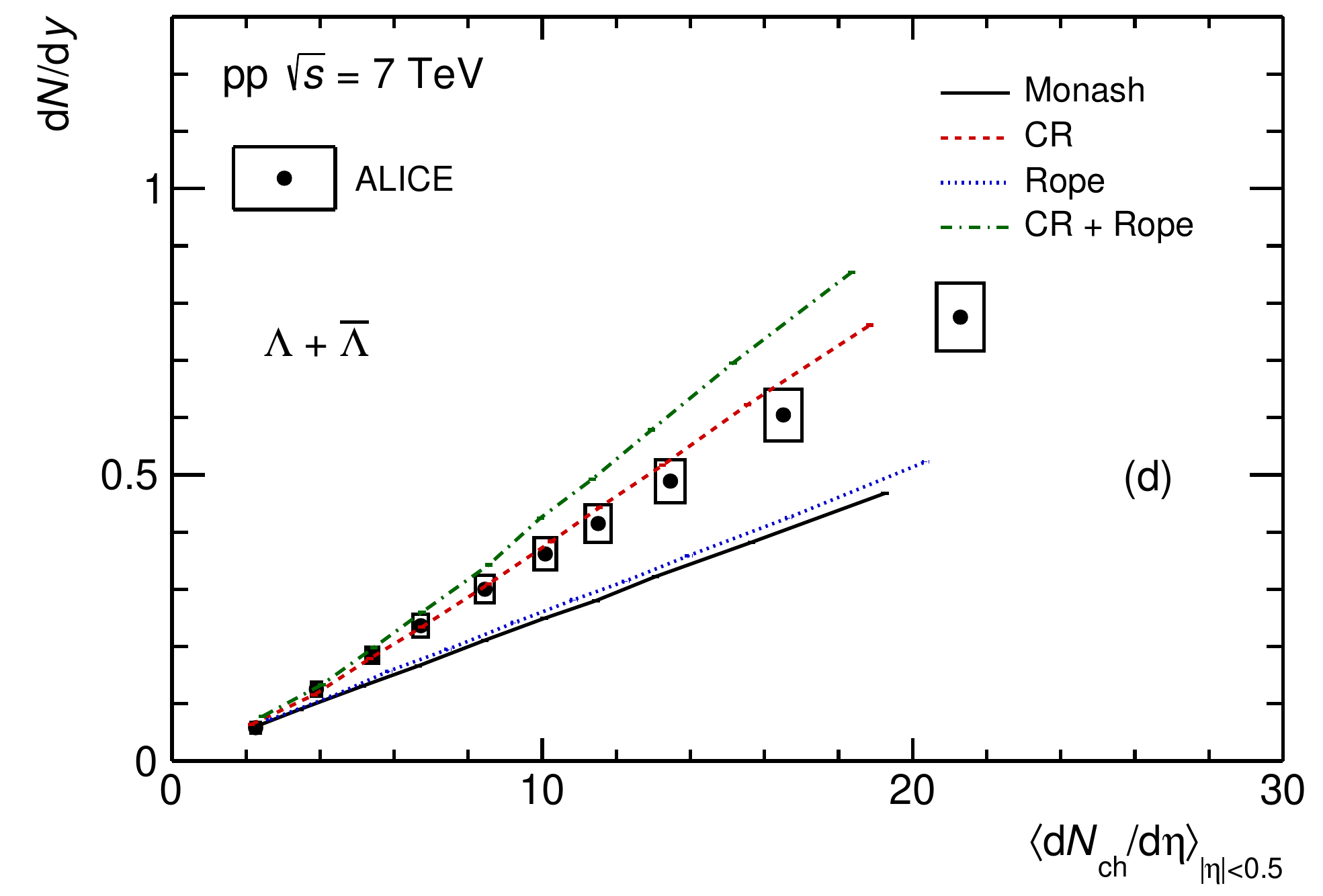}
		\includegraphics[width=.32\textwidth]{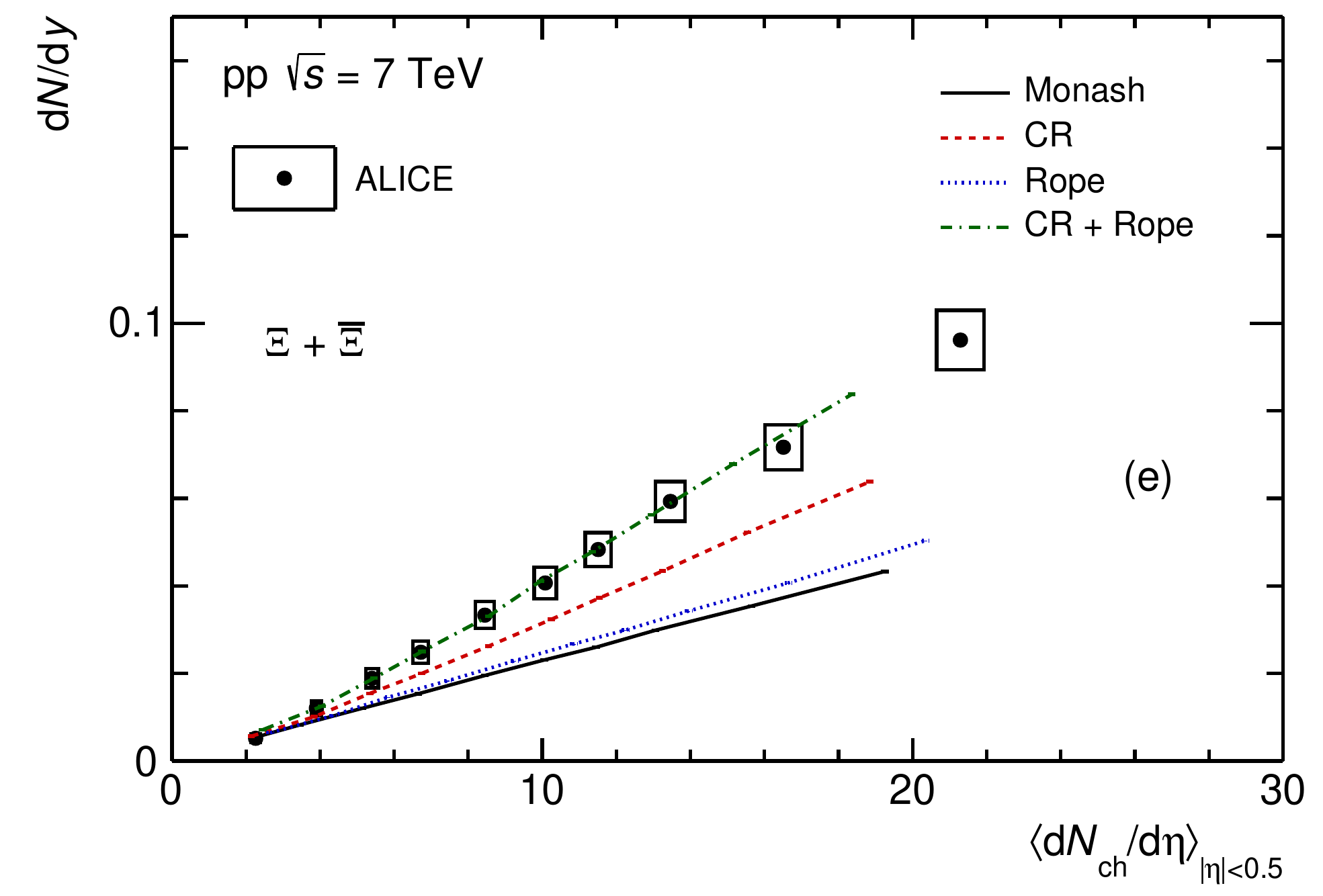}
		\includegraphics[width=.32\textwidth]{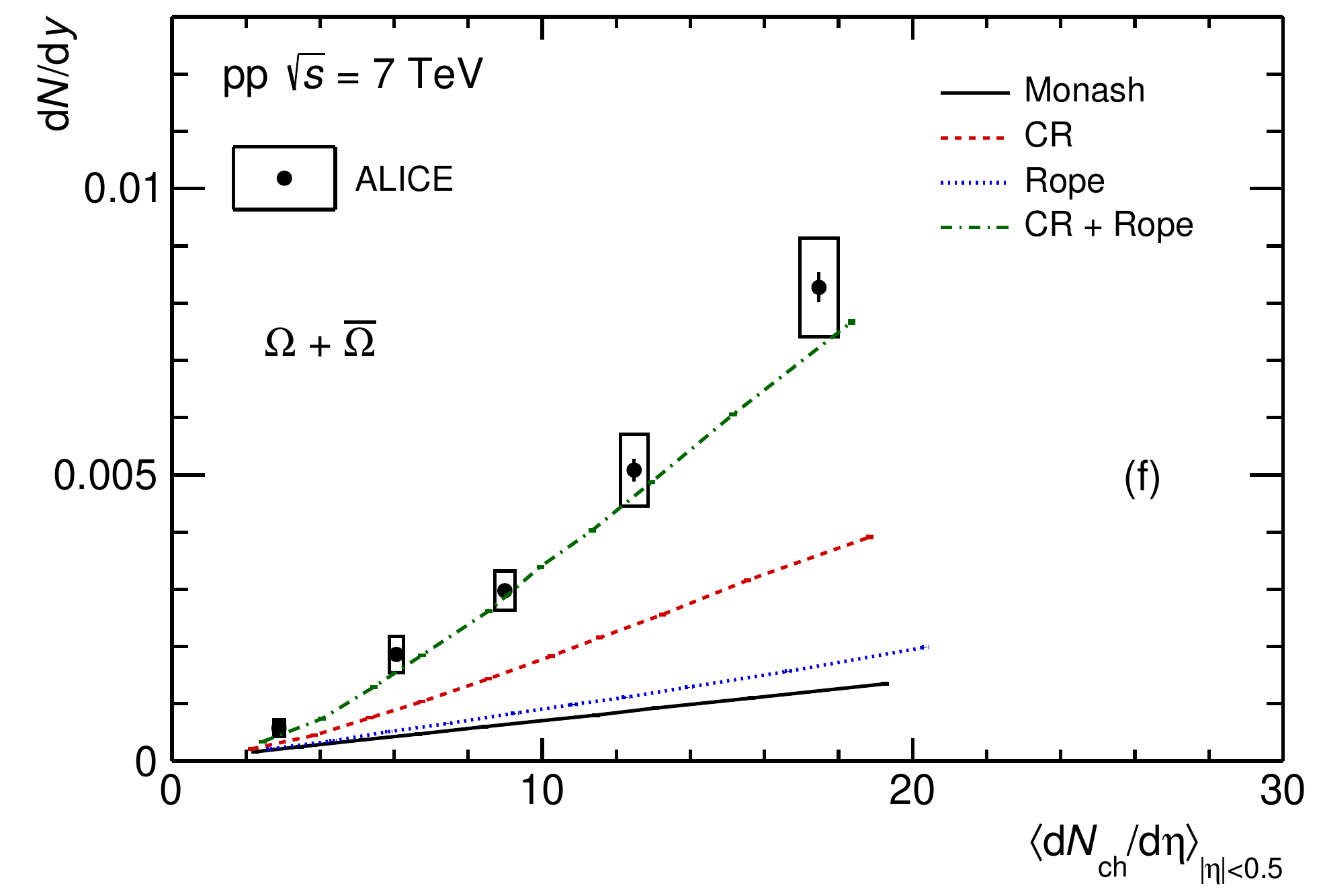}
	\end{center}
	\caption{Integrated yields d$N/$d$y$ of various hadrons, $\kzero$, $\kstar$, $\phi$, $\lmb$, $\XI$, and $\OMS$, as functions of $\langle\dndeta\rangle_{|\eta| < 0.5}$. The mesons yields are shown in the top plots, and the baryons yields are shown in the bottom plots. Model results are show for \pp collisions at \seven, data for \pp collisions at \seven and \thirteen (only for $\phi$ particle). The data point are taken from \cite{ALICE:2016fzo, ALICE:2019etb}.}
	\label{fig:InclIntePar}
\end{figure*}

The integrated yields of \kzero, \kstar, \PHI, \lmb, \Xis, and \Oms in \pp collisions at \seven varying with the event multiplicity are shown in Fig.~\ref{fig:InclIntePar}. The Monash reference results are shown with the black solid lines. Color reconnection and rope model effects are represented by the red dashed lines and the blue dotted lines, while the  CR and rope combined effects are presented in the green dash-dotted lines. It is observed that the CR effects generally enhance the productions in the baryon sector and suppress the meson productions due to the creation of additional junction structures at the beyond the leading color level. The rope model, however, slightly increases the yield of all hadrons. Combining the CR and rope model, the meson suppression due to CR is compensated and the enhancement to the baryon production becomes stronger in the multi-strange particles as an outcome of the string tension changes. 

\begin{figure*}[ht]
	\begin{center}
		\includegraphics[width=.32\textwidth]{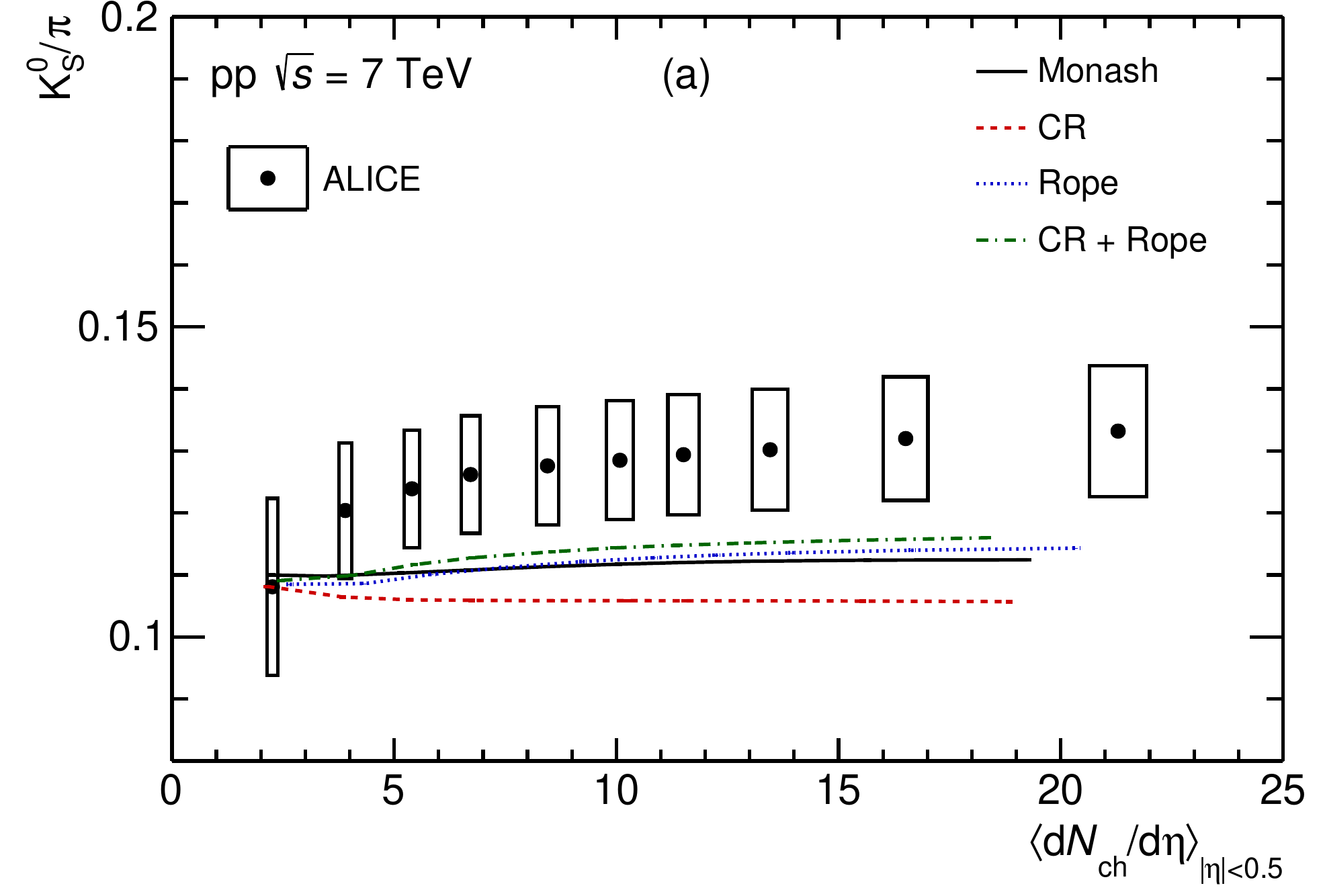}
		\includegraphics[width=.32\textwidth]{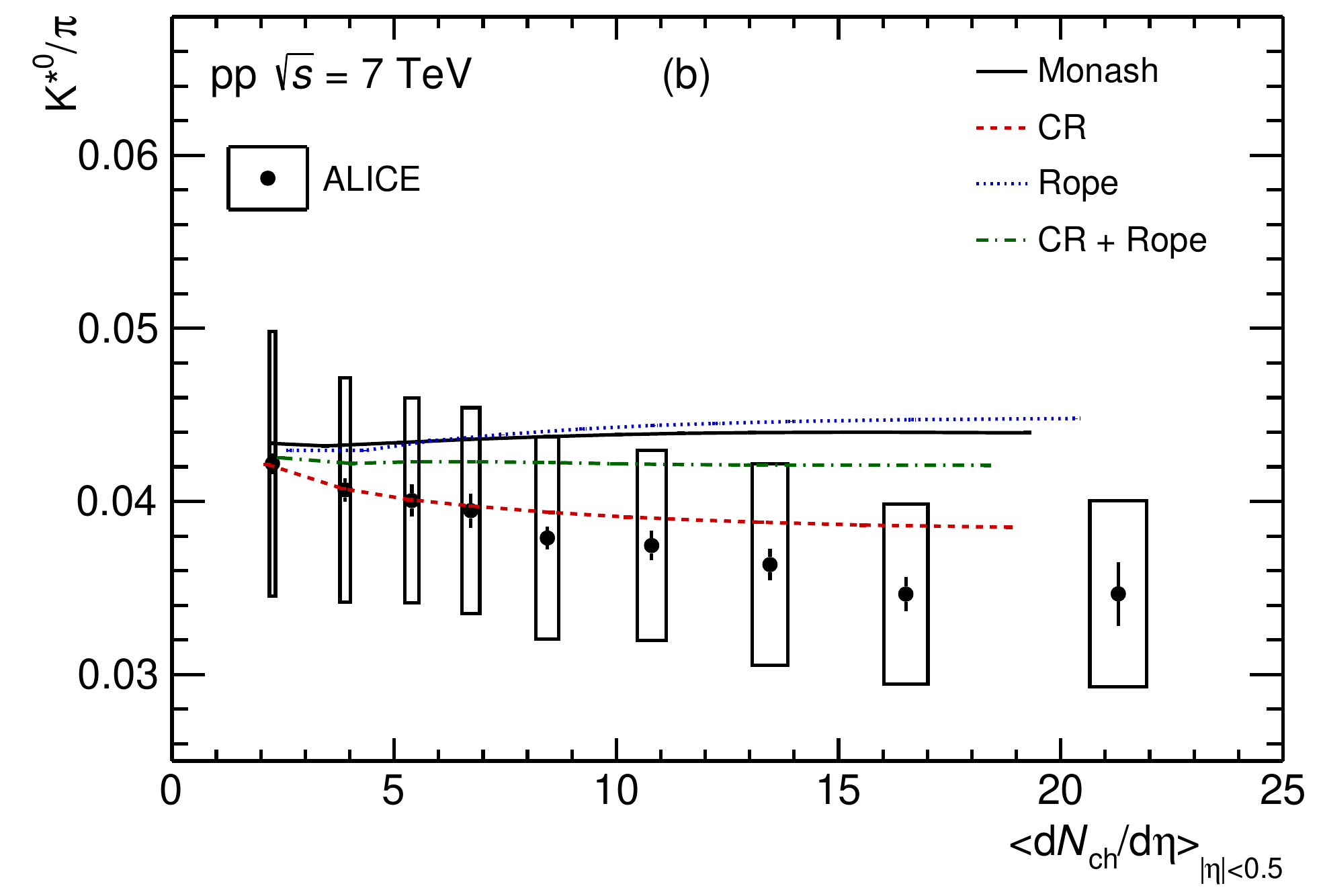}
		\includegraphics[width=.32\textwidth]{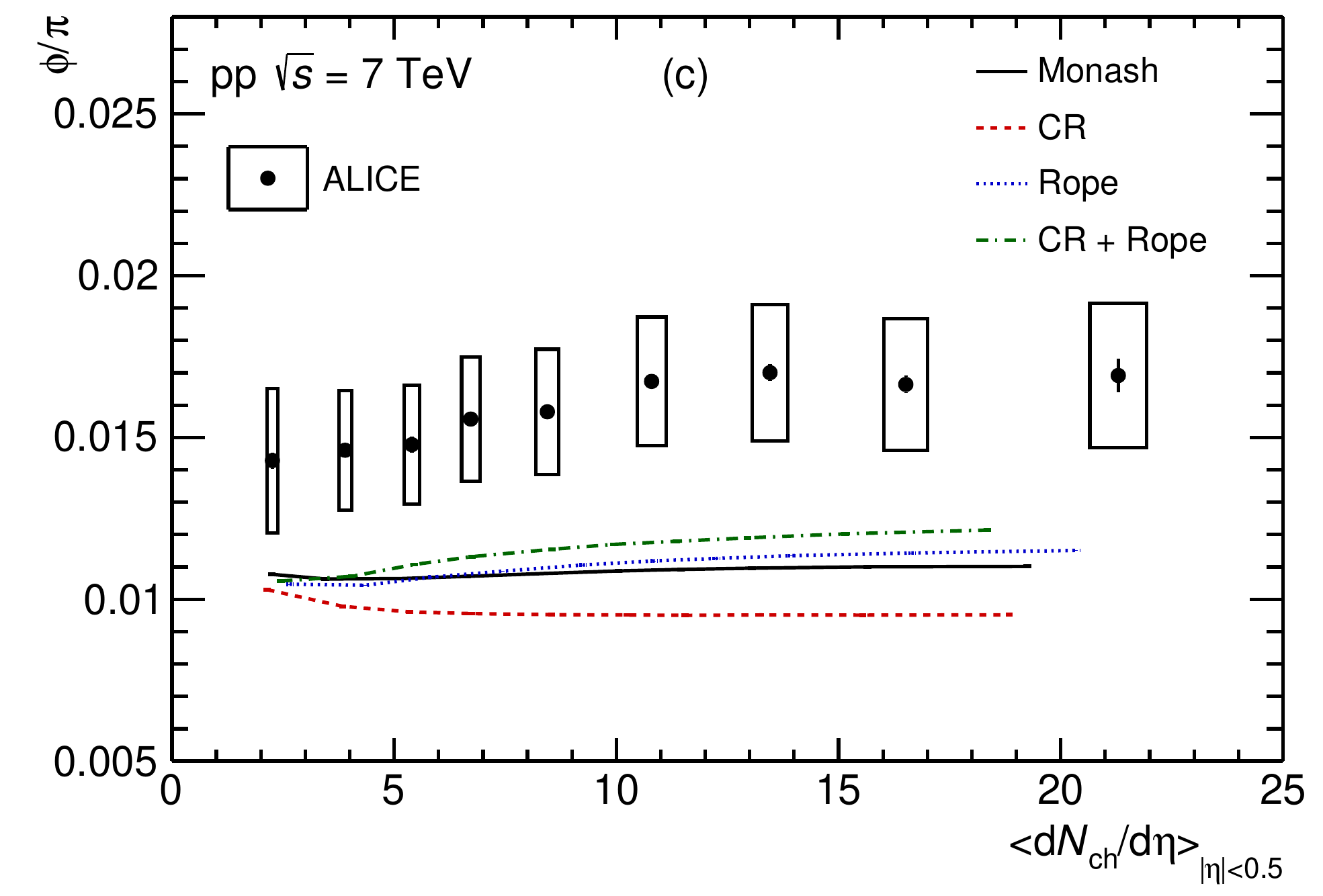}
		\includegraphics[width=.32\textwidth]{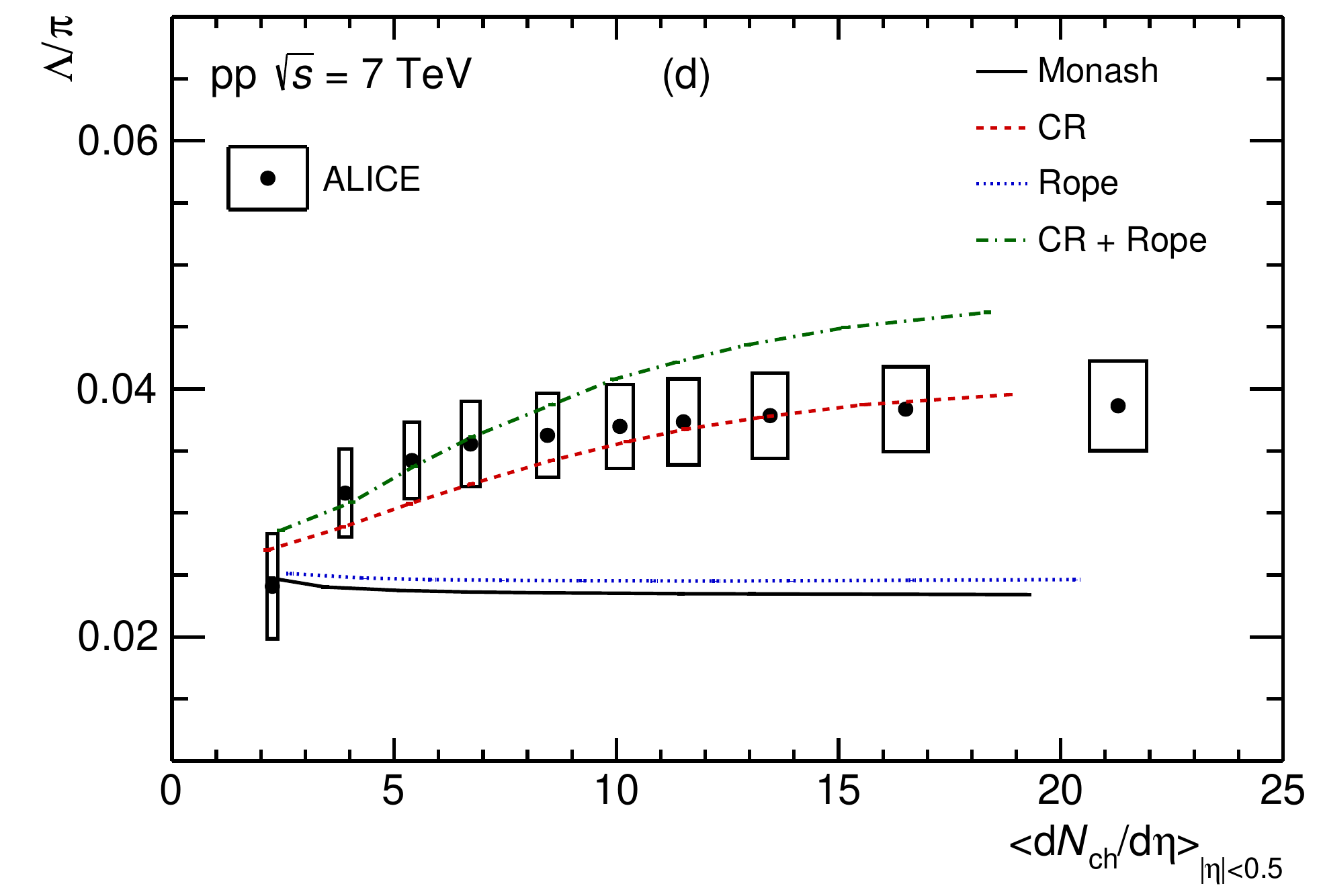}
		\includegraphics[width=.32\textwidth]{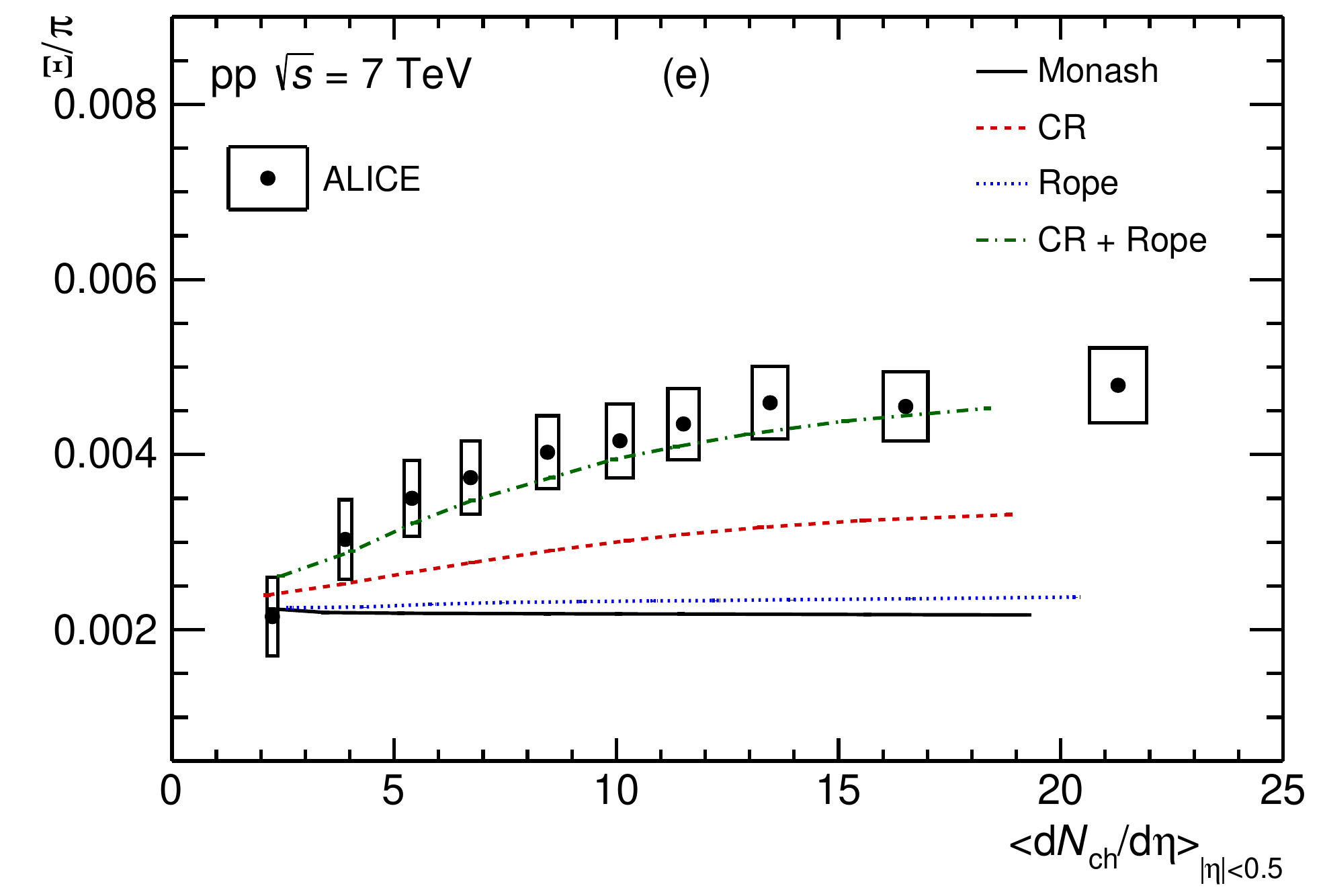}
		\includegraphics[width=.32\textwidth]{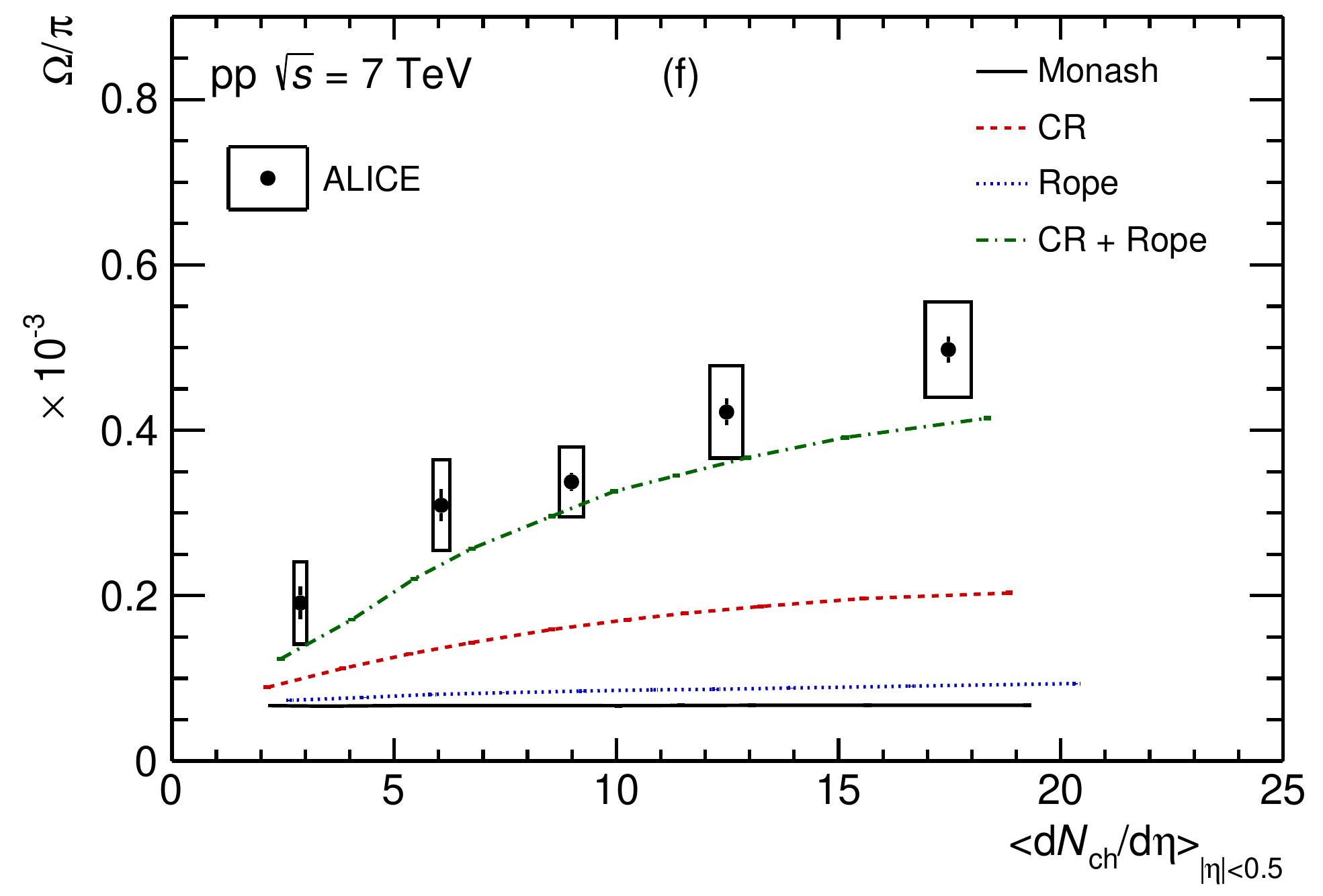}
	\end{center}
	\caption{Hadrons to $\pi$ ratios, $\kzero/\pi$, $\PHI/\pi$, $\lmb/\pi$, $\XI/\pi$, and $\OMS/\pi$, as function of $\langle\dndeta\rangle_{|\eta| < 0.5}$ distributions in \pp collisions at \seven.  The mesons to $\pi$ ratios are shown in the top plots, and the baryons to $\pi$ ratios are shown in the bottom plots. The data point are taken from  \cite{ALICE:2016fzo, ALICE:2018pal}.}
	\label{fig:InclIntePartoPiRatio}
\end{figure*}

The multiplicity dependent distributions of all the strange particle to pion ratios are presented in Fig.~\ref{fig:InclIntePartoPiRatio}. It is clearly seen that the baryon to meson ratio increases with the event multiplicity in the CR model as shown by the red dashed lines, resulted from the fact that more non-leading connected partons can form junctions in the high multiplicity events. The resonance meson states \kstar and \PHI to pion ratios are gradually decreasing with \dndeta, consistent with the previous observation that lower mass states are preferred in the breakup of strings with shorter length in the color reconnection model~\cite{Goswami:2019mta, Acconcia:2017bjv}. After including the baryon enhancement from CR, the CR combined rope model reasonably describes especially the multi-strange baryon to meson ratio, which is an important signature of the strangeness enhancement. The rapid growth of the \Xis and \Oms to pion ratio with \dndeta is a result of the large string tension of color rope fragmentations in the high multiplicity events. Strange quarks and spin 1 $ss$ diquarks are more likely to be produced with larger string tensions during the string breakups in the rope hadronization model. It is also noted that the resonance production feature in the rope model is different from CR effects. \kstar and \PHI meson to pion ratios are both enhanced in high multiplicity events with only the color rope mechanism. The combined model, however, brings down the \kstar yield while the \PHI yield is further enhanced. To sum up, the color rope model combined with CR effects is expected to provide the best inclusive strange hadron production estimations. 

\subsection{Strange particle production in jets and UE}
\label{subsec:jetres}

\begin{figure*}[ht]
    \centering
    \includegraphics[width=.49\textwidth]{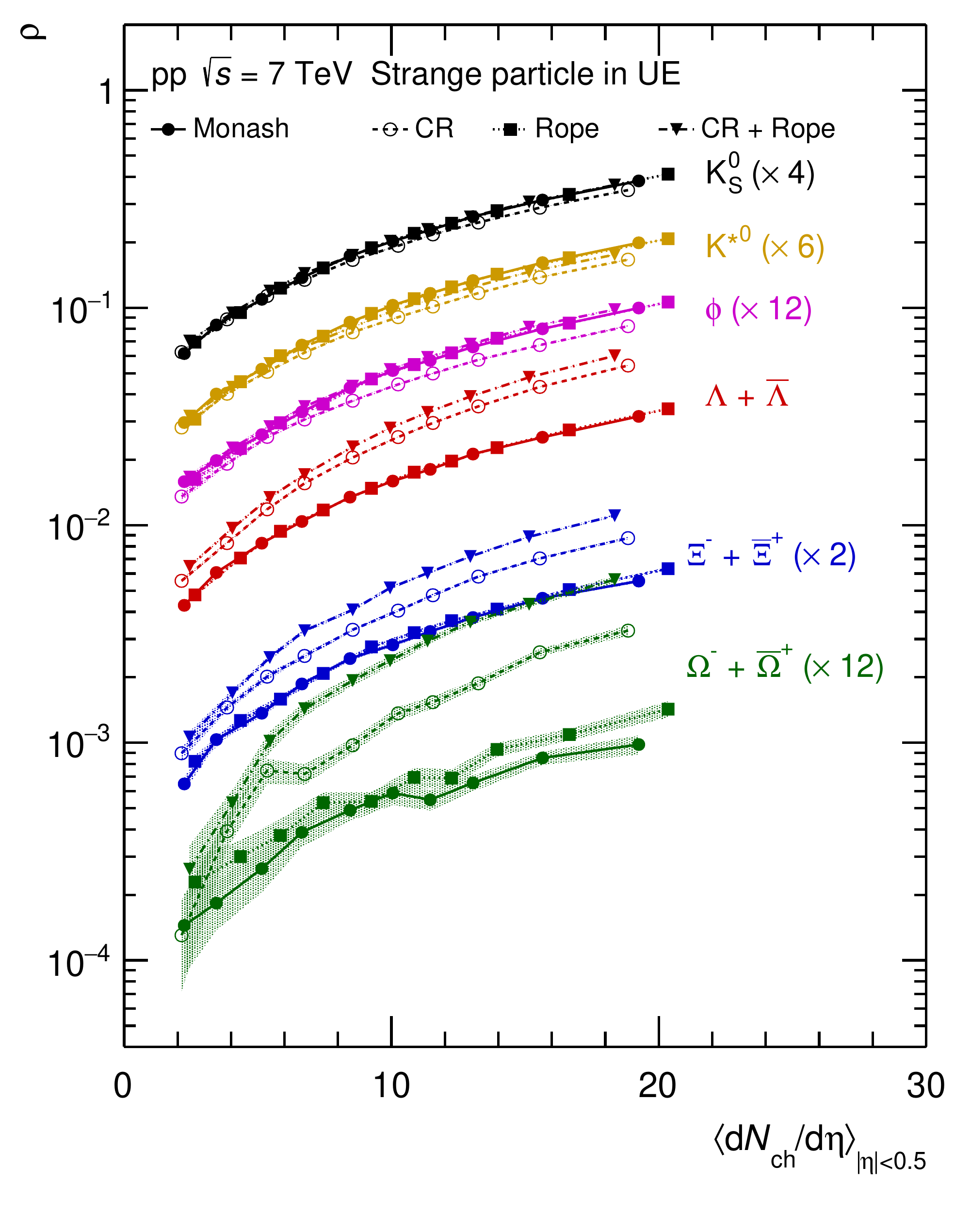}
    \includegraphics[width=.49\textwidth]{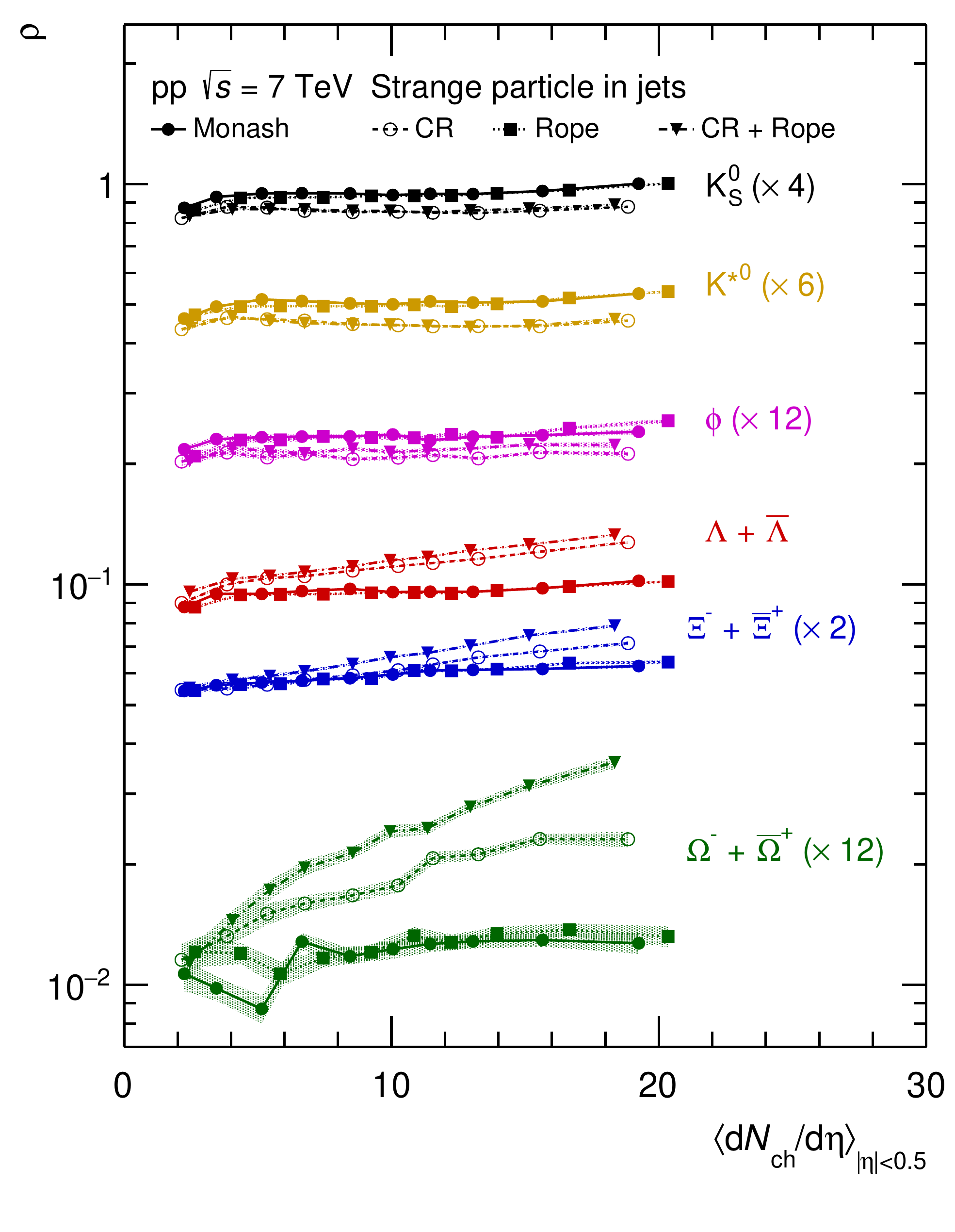}
    \caption{The integrated yields of \kzero, \kstar, $\phi$, \lmb, $\XI$, and $\OMS$ in underlying events (left) and in charged-particle jets (right) as a function of $\avdndeta_{|\eta|<0.5}$ distributions in \pp collisions at \seven.}
    \label{fig:UEJEIntePar}
\end{figure*}
In this section, we study the strange hadron productions associated with hard processes with the help of jet measurements. We estimate the strange hadron yield from soft processes by investigating the hadron productions perpendicular to the jet cone direction. The strange hadron yields normalized to per unit acceptance area $\rho$ in PC as a function of $\avdndeta_{|\eta| < 0.5}$ are shown in the left panel of Fig.~\ref{fig:UEJEIntePar}. Similar to what has been shown in Fig.~\ref{fig:InclIntePar}, the CR induced meson suppression and baryon enhancement persists in the PC hadron productions. The strangeness number dependent baryon enhancement is also observed in the CR and rope combined model. It can be inferred from this resemblance that the inclusive strange hadron production is largely dominated by the UE effects.  

The in-jet integrated yields of \kzero, \kstar, $\phi$, \lmb, $\XI$, and $\OMS$ as a function of $\avdndeta_{|\eta| < 0.5}$ are presented in the right panel of Fig.~\ref{fig:UEJEIntePar}. It is distinguishing that the strange hadron productions in PC are largely driven by the event multiplicity while the in-jet strange meson yields are quite stable over a wide multiplicity range. This is not unexpected as the hadron production in UE is strongly related to the number of multiple parton interactions in an event but the in-jet hadron yield is more sensitive to the jet fragmentation effects. Therefore, the in-jet meson yield only slightly changes with the event charged-partice multiplicity density since the per jet energy, which dictates the in-jet hadron productions, is hardly changed at different event activity. On the other hand, the CR effect with or without color rope mechanism leads to significant increase of the baryon yield inside the jet cone as a function of the event multiplicity, indicating the CR effect is important for both soft and hard processes. The faster increasing trend of the baryon production inside jets also suggests that the junction structure due to color reconnection is more likely to be formed inside jets. This is consistent with the picture that beyond leading color reconnections often happen between different MPI initiated string ends, thus quite impressive to the high $\pT$ physics.  


\begin{figure*}[!h]
    \centering
    \includegraphics[width=.49\textwidth]{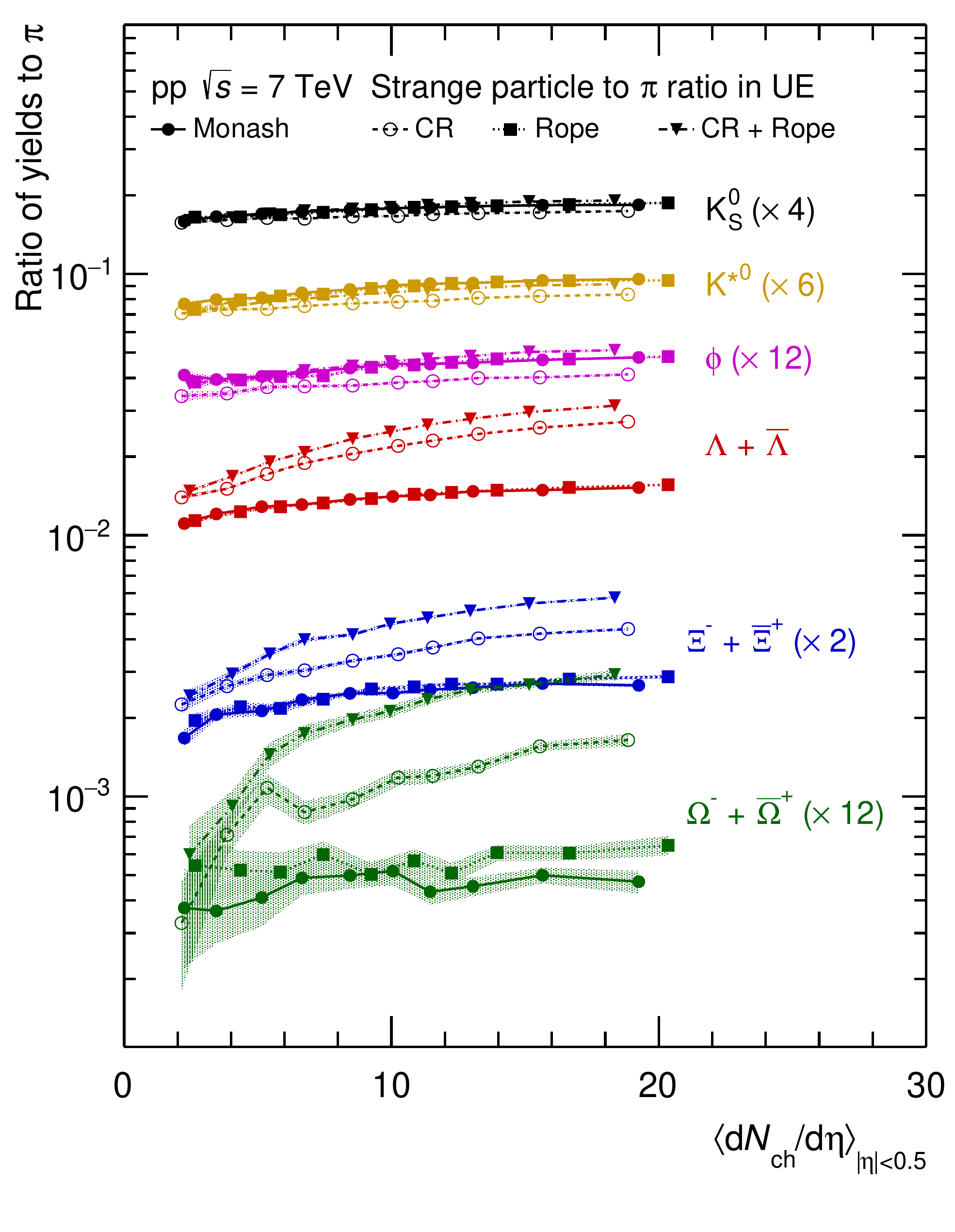}
    \includegraphics[width=.49\textwidth]{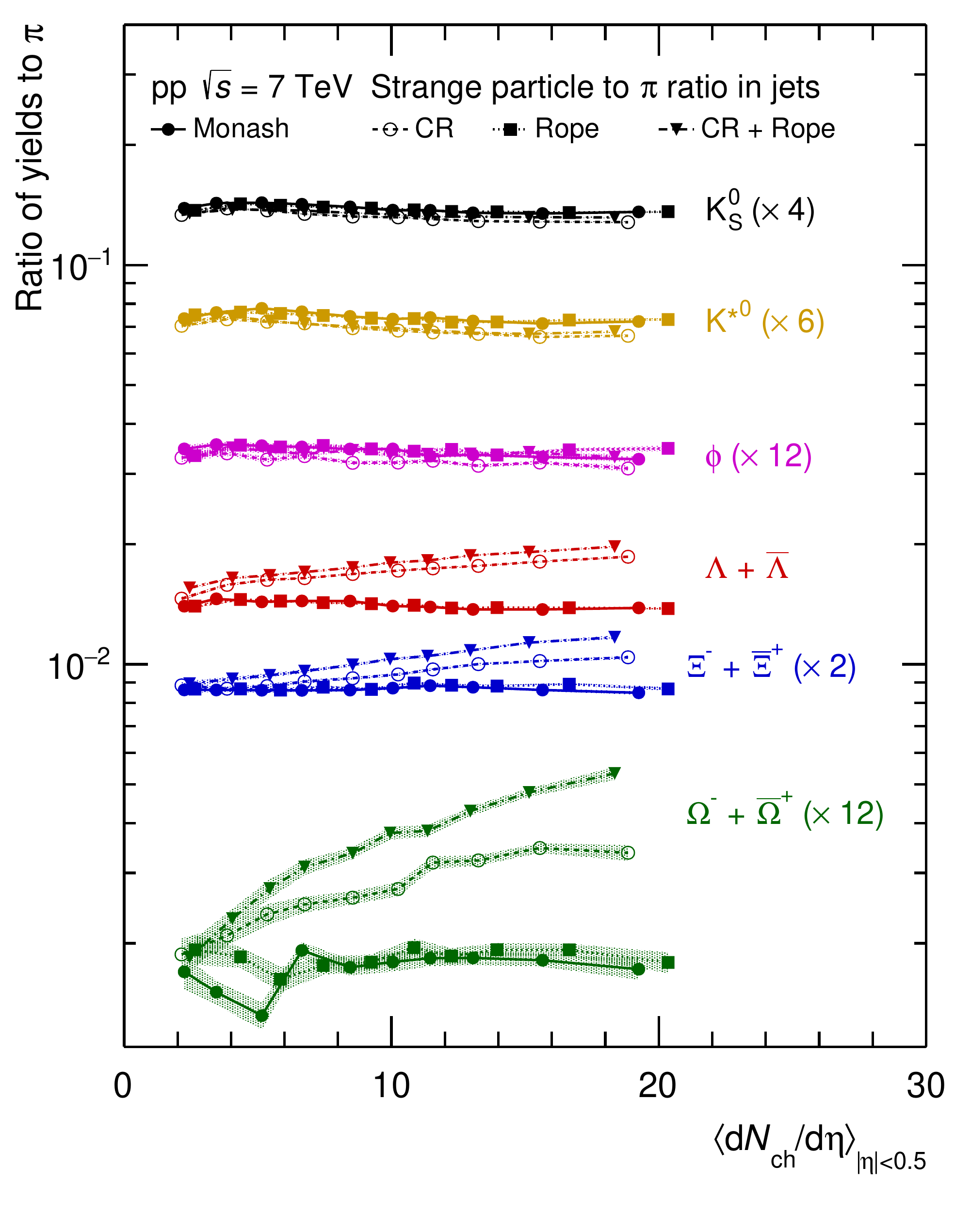}
    \caption{The integrated yield ratios of strange hadrons to $\pi$ in underlying events (left) and in charged-particle jets (right) as a function of $\avdndeta_{|\eta|<0.5}$ distributions in \pp collisions at \seven.}
    \label{fig:UEJEIntePartoPiRatio}
\end{figure*}

The strange hadron to $\pi$ ratios,  $\kzero/\pi$, $\phi/\pi$, $\lmb/\pi$, $\XI/\pi$, and $\OMS/\pi$ as functions of $\avdndeta_{|\eta| < 0.5}$ distributions in underlying evens of \pp collisions at \seven derived by the integrated yields are shown in the left panel of Fig.~\ref{fig:UEJEIntePartoPiRatio}. The similar increasing trend compared to Fig.~\ref{fig:InclIntePartoPiRatio} confirms that the event activity dependence of the strange to non-strange particle ratio observed in the inclusive distribution is from soft processes. 

The in-jet strange hadron to pion ratios in \pp collisions at \seven varying with $\avdndeta_{|\eta| < 0.5}$ are presented in the right panel of Fig.~\ref{fig:UEJEIntePartoPiRatio}.  It is found that the strange meson to pion ratios including those for the strange resonance in jets are almost independent of the $\avdndeta_{|\eta| < 0.5}$, suggesting the high $\pT$ parton flavor is more likely to be determined by the hard QCD process and not sensitive to the fragmentation details. The striking increase of the in-jet baryon yield is manifested in the strange baryon to pion ratio as well. The rope model combined with CR effect delivers the strangeness number dependent baryon enhancement feature like that found in the inclusive distributions. It is also observed that unlike the strange baryon to pion ratios in the inclusive distributions and PC, which saturate at the intermediate \dndeta, the in-jet strange baryon to pion ratios grow faster in the high multiplicity events. 

\begin{figure*}[ht]
    \centering
    \includegraphics[width=.48\textwidth]{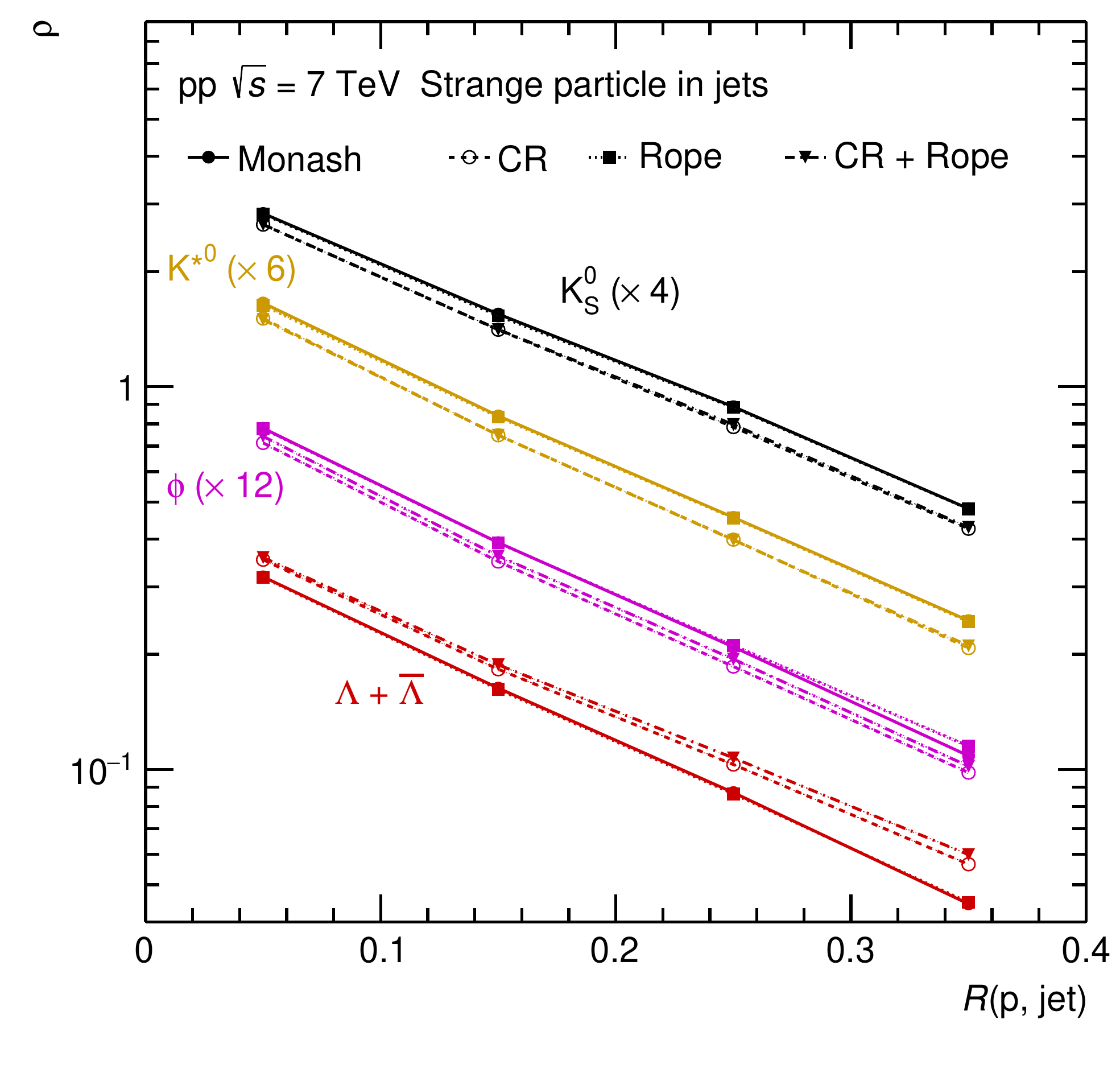}
    \includegraphics[width=.48\textwidth]{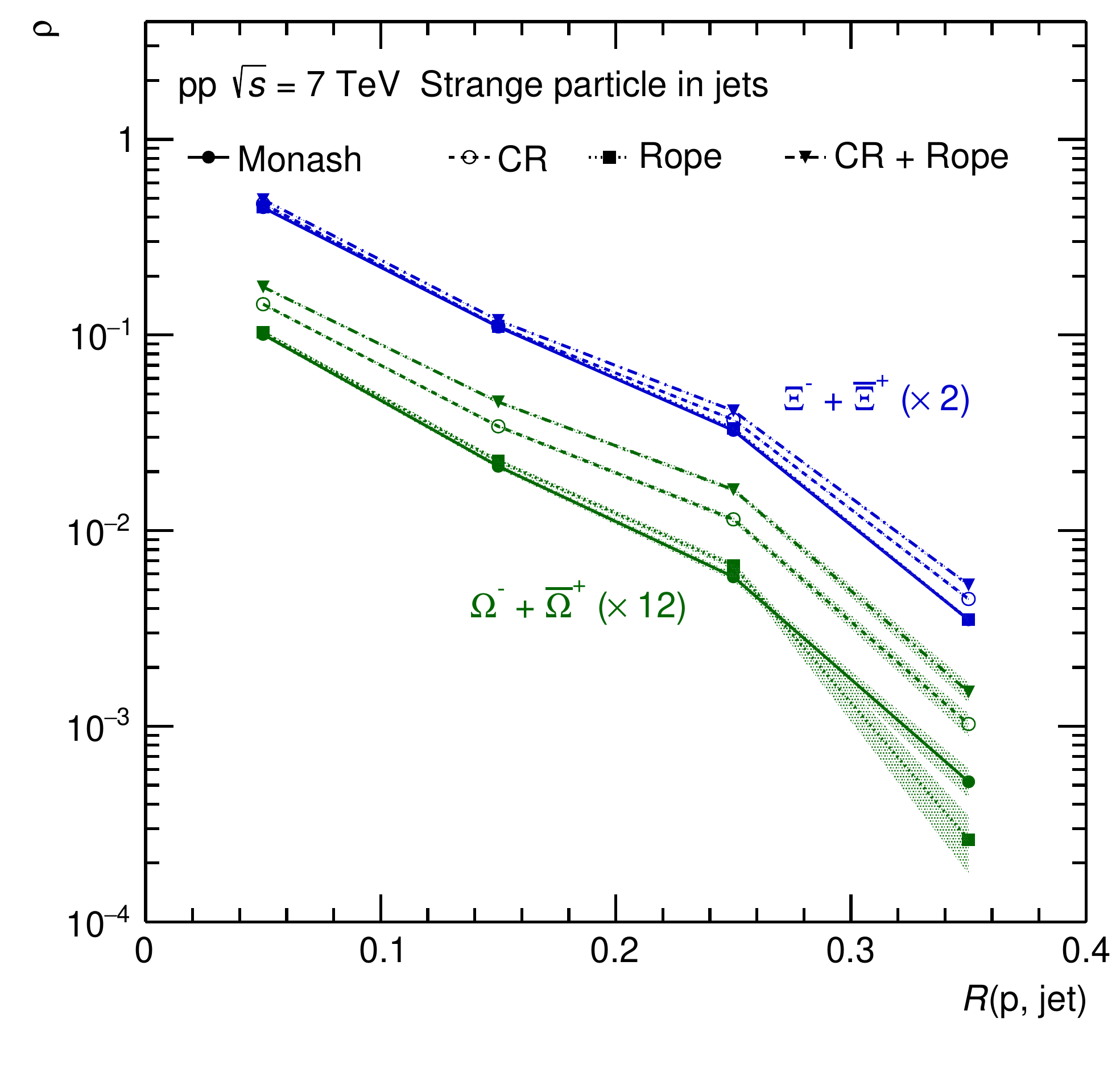}
    \caption{The integrated yields of \kzero, \kstar, \PHI, \lmb (left) and \XI, and \OMS (right) in jet cone as a function of $R$ (p, jet).}
    \label{fig:PartoJet}
\end{figure*}

The jet shape dependence of the strange hadron yields is also investigated. In Fig.~\ref{fig:PartoJet}. The yields of \kzero, \kstar, \PHI, \lmb, \XI and \OMS as functions of the distance between particle to the jet axis $R$(p, jet) are presented. It is found that the decreasing trend of \kzero, \kstar, \PHI and \lmb strange hadron with $R$(p, jet) are almost the same. A faster decreasing is observed for multi-strange baryons, \XI and \OMS, at large $R$(p, jet).
\begin{figure*}[ht]
    \centering
    \includegraphics[width=.48\textwidth]{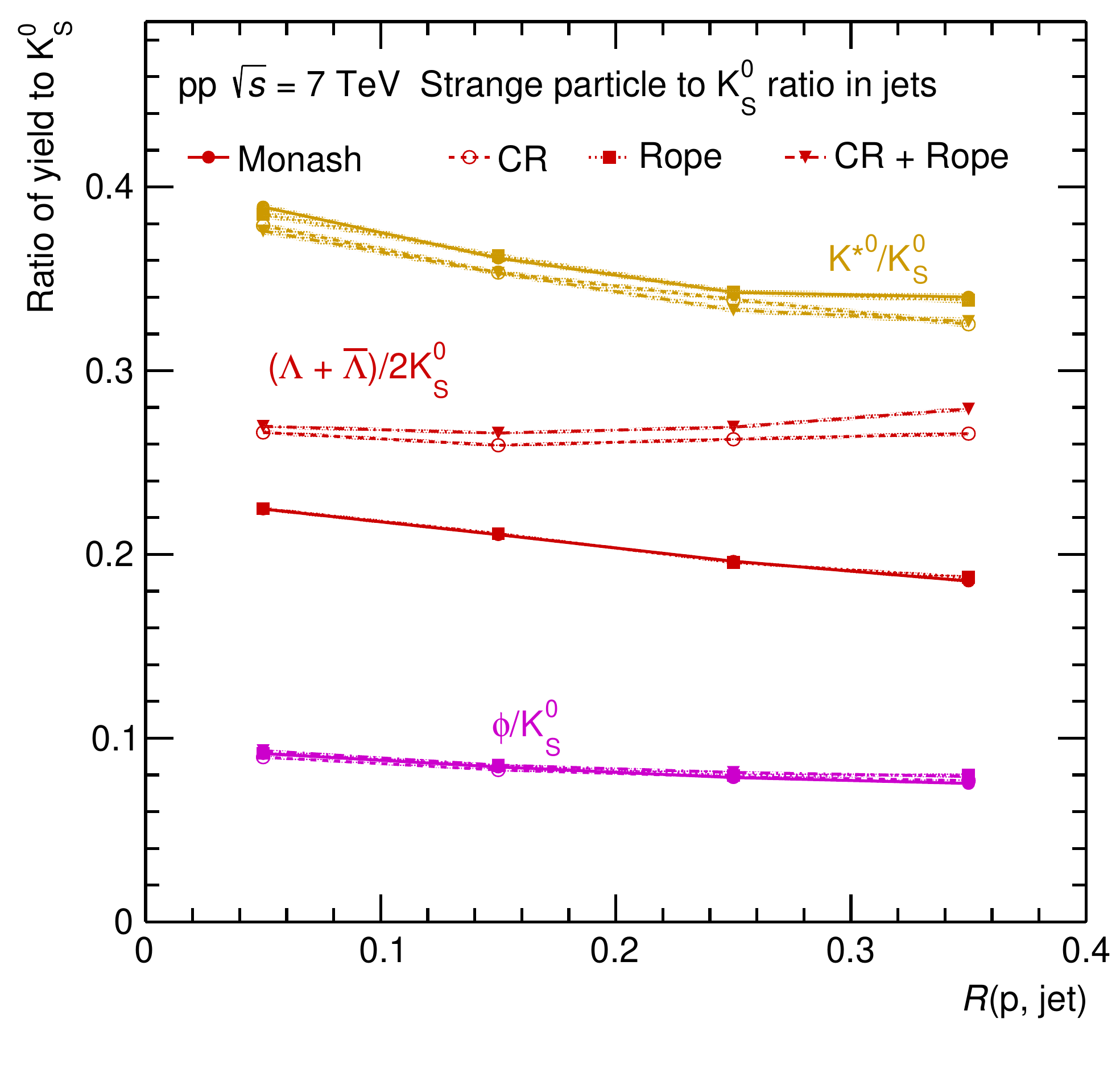}
    \includegraphics[width=.48\textwidth]{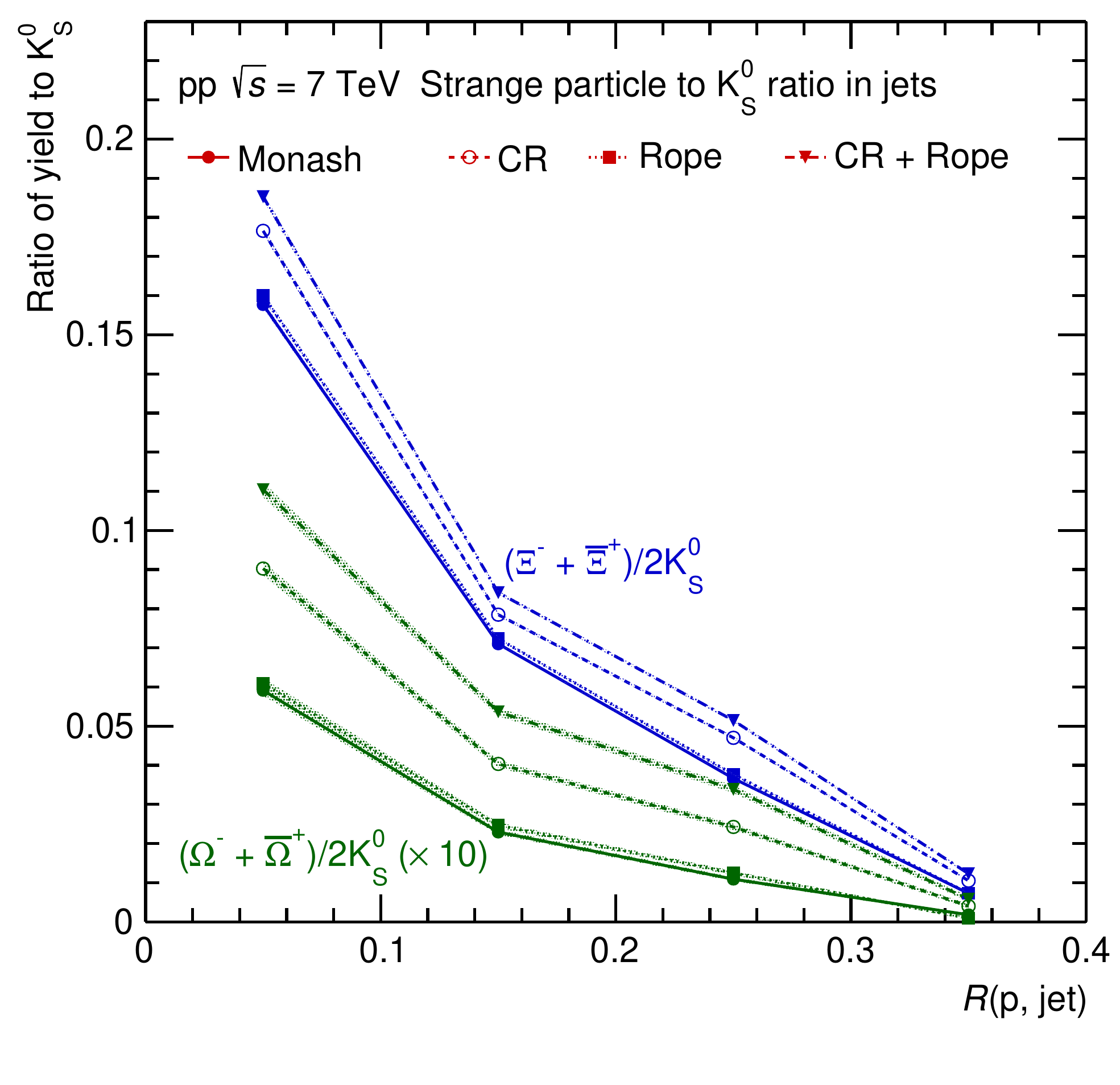}
    \caption{The ratios of \kstar/\kzero, $\PHI$/\kzero, \lmb/\kzero (left) and $\XI$/\kzero, and $\OMS$/\kzero (right) in jet cone as a function of $R$ (p, jet).}
    \label{fig:ParRtoJet}
\end{figure*}

We also investigate the jet shape dependence of the relative strange hadron yield ratios in Fig.~\ref{fig:ParRtoJet}. In this comparison, the ratios of \kstar, $\PHI$, \lmb, $\XI$, and $\OMS$ to \kzero in jet cone as function of distance between particle to jet axis $R$(p, jet) are presented. It is found in the Monash reference all the strange hadron to \kzero ratios are decreasing with $R$(p, jet). The slow variation of strange resonances and \lmb to \kzero ratios implies that the in-jet strange resonance mesons and \lmb baryons are uniformly produced along the string piece similar to \kzero. The more pronouncing $R$(p, jet) dependence for the multi-strange baryons indicates that multi-strange baryon formations in the string fragmentation framework are strongly related to the junction structures created at the string ends. The multi-strange baryons are thus more likely to be produced close to the jet axis, possibly being the leading hadron inside the jet. It is interesting to see that the CR model provides different behaviors for \lmb to \kzero and multi-strange baryon to \kzero ratios. This feature can result from that \lmb baryons can be created with the reconnected junctions formed between the beam remnant parts, thus pulling the \lmb baryons away from the jet axis direction close to the high $\pT$ side of the string end. The enhancement of multi-strange baryons in the CR model comes from the connection of two already formed $s$ quarks, which are expected to be associated with the hard production, leading to the enhanced production of strange baryons close to the jet axis direction. This feature suggests the in-jet multi-strange baryon productions can be potentially unique probes to the hard QCD process related physics in the string fragmentation picture.

\section{Summary}
\label{sec:sum}
In summary, the production of strange hadrons is studied in \pp collisions at \seven with the modified string fragmentation models within the PYTHIA8 framework. The inclusive integrated yields and relative production ratios to pion of strange particles as functions of the event multiplicity can be well described by the combined color reconnection and color rope effects. The CR model generally enhances the productions in the baryon sector and suppress the meson productions with respect to Monash tune. The strange hadron productions are also studied in events with energetic jets by dividing the particles into the in-jet region and the underlying event region which are expected to represent the hard and soft process effects, respectively. The results indicate the inclusive strange particle production is dominated by the underlying event contributions. The multiplicity dependence of the strange particle yield in jets is found to be much weaker compared to the rapid growth in underlying events, suggesting that the in-jet hadron production is more sensitive to the jet fragmentation process. However, the color reconnection and color rope induced enhancement of multi-strange hadron to pion ratio is visible both in the soft and the hard process. The untamed increasing of the in-jet strange baryon yield implies the strange baryon productions are largely related to the hard process. The in-jet yields of strange particles are also dependent on their distance to the jet axis. The multi-strange baryons are more likely to be produced close to the jet axis compared to the strange mesons. These observations may shed more light on the understanding to the multiplicity dependent strangeness enhancement in pp collisions at LHC energies.

\begin{acknowledgements}

This work is supported by the National Key Research and Development Program of China (2016YFE0100900), the National Natural Science Foundation of China (11875143, 11905188 and 12061141008) and the Innovation Fund of Key Laboratory of Quark and Lepton Physics LPL2020P01 (LZ).

At the end we would like to thank Christian Bierlich for providing us the input parameters for the color reconnection and the color rope model.
\end{acknowledgements}

\section*{Appendix}
To clarify the reliability of the treatment to the UE contribution in this work, we provide in the Appendix additional details for the jet number distributions in different multiplicity bins, which explores the multi-jet impacts on our analysis.
\begin{figure}[!h]
	\begin{center}
	    \includegraphics[width=.48\textwidth]{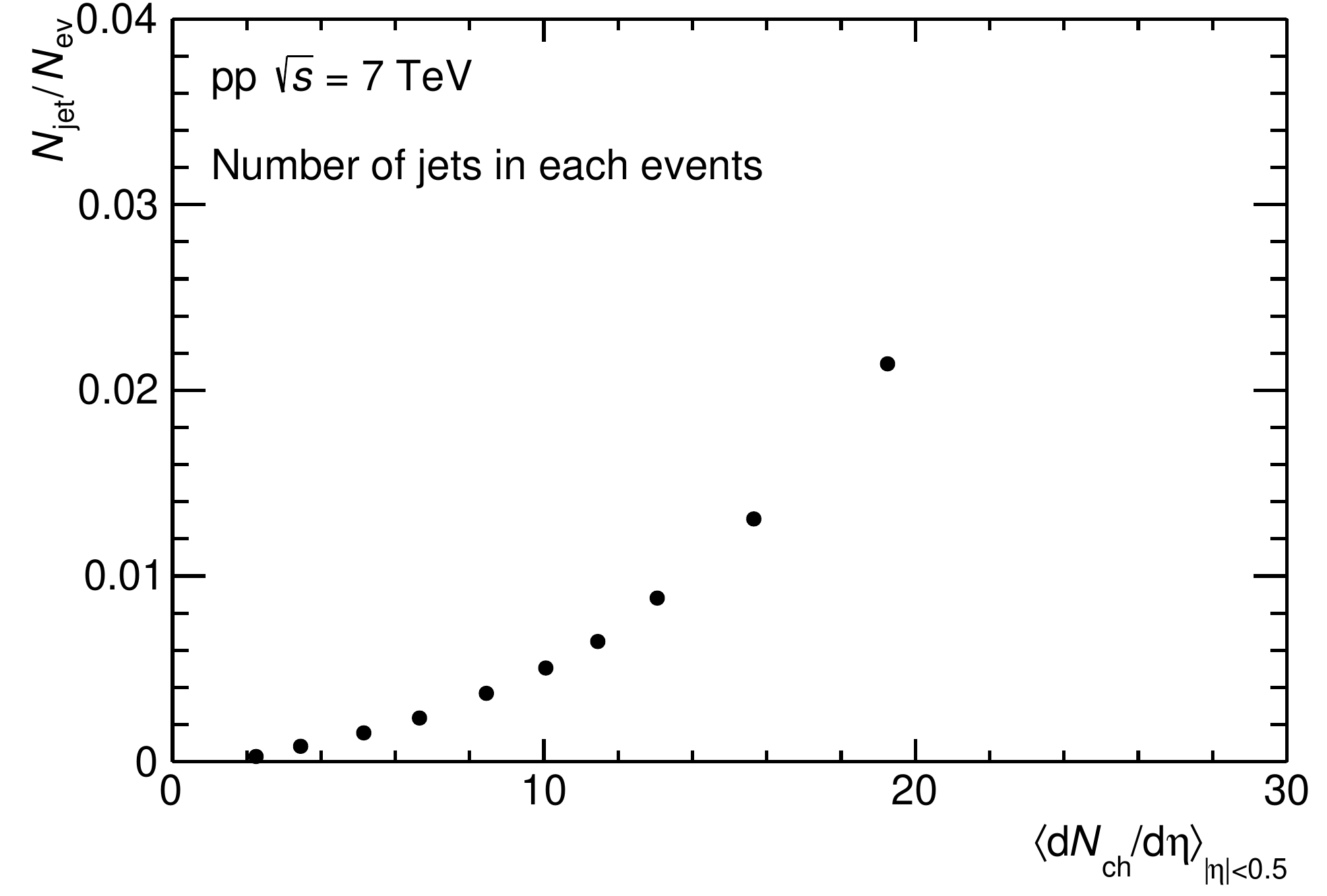}
	    \includegraphics[width=.48\textwidth]{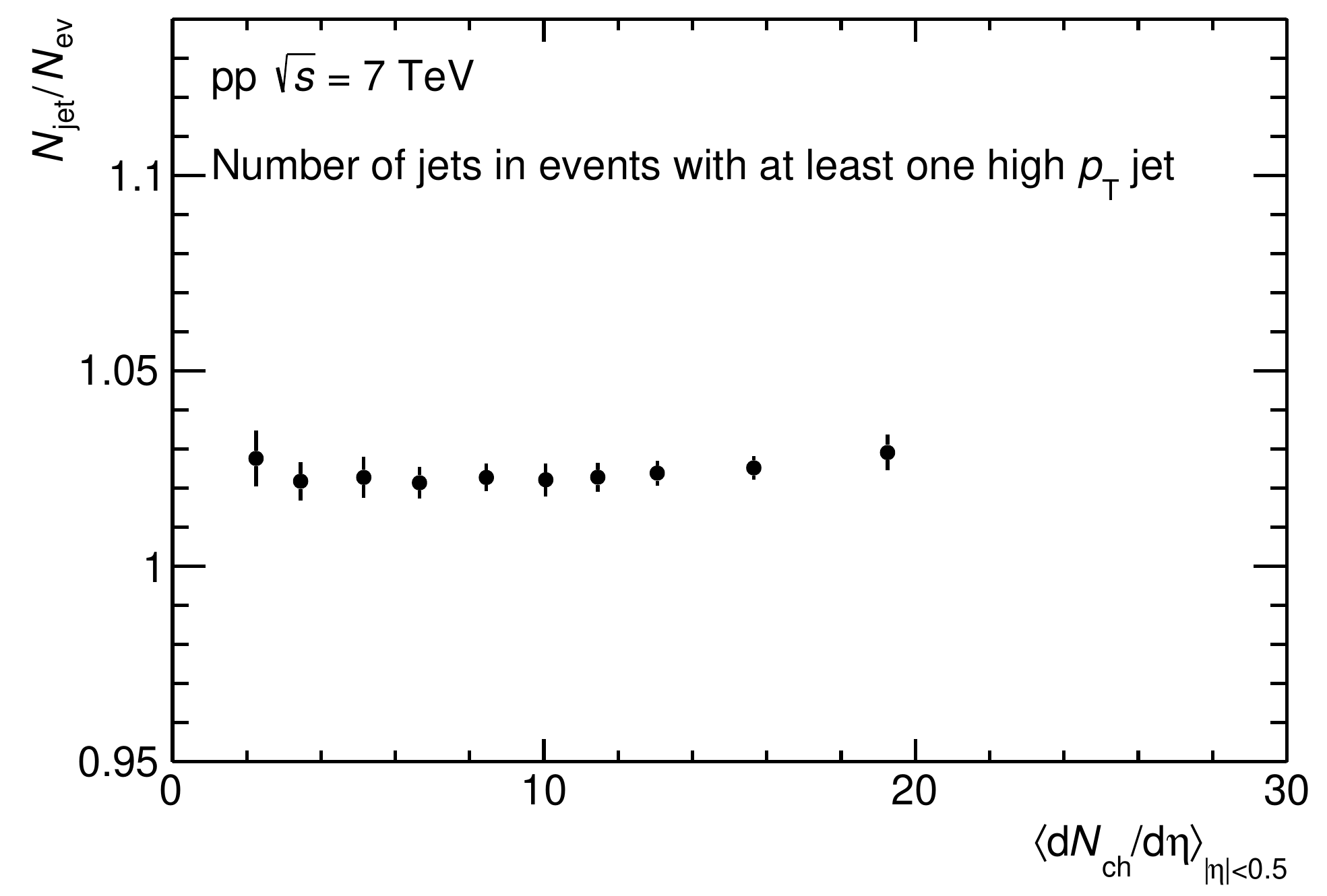}
		\includegraphics[width=.48\textwidth]{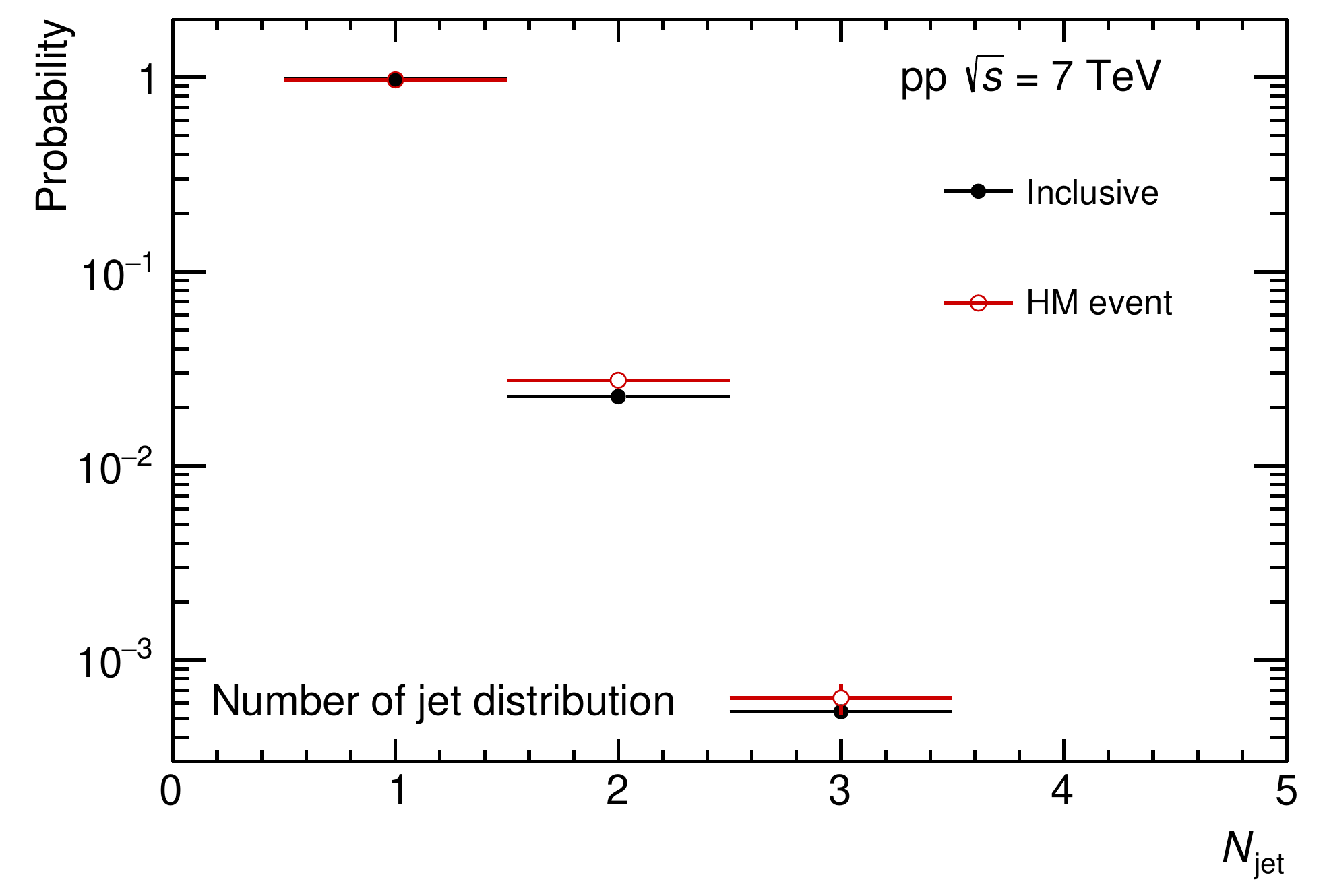}
	\end{center}
	\caption{Jet number distribution varying with the event multiplicity in each minimum bias events (top), jet number with the event multiplicity distribution in events with at least one high $\pT$ jet (middle), and jet number probability distribution in the most high multiplicity event class (bottom) compared to that in inclusive events.}
	\label{fig:appendix}
\end{figure}

The top panel of Fig.~\ref{fig:appendix} presents the yields of $\pTjch > 10$~\GeVc jets dependent on the event multiplicity from the default PYTHIA8 simulations. The jet yields grow rapidly with the event multiplicity, while the per-event high $\pT$ jet number is around 0.02 even in the highest multiplicity bin. We also explore the multiplicity dependence of jet yields normalized to the events with at least one high $\pT$ jet in the middle panel of Fig.~\ref{fig:appendix}. It is shown that the produced jet number in each jet event is very close to one and almost independent of the event activity. The jet number probability distribution in high multiplicity jet events is found to be similar to that in the minbias jet events as shown in the bottom panel of Fig.~\ref{fig:appendix}. Most of the events contain only one jet and the probability to have more than one back-to-back di-jet in an event is very low. In that sense, the contamination to UE from jet particles is negligible within the accessible multiplicity range which can be covered by the current experiments in pp collisions at the LHC energy. Like what has been done in many jet related experimental studies~\cite{ALICE:2021cvd, ALICE:2019mmy}, the UE contribution can be safely estimated simply by taking the perpendicular region to the leading jet direction, which is denoted as the perpendicular cone method in this work.

\bibliographystyle{spphys}       
\bibliography{StrinJet}   

%
%
\end{sloppypar}
\end{document}